\documentclass[10pt,twocolumn,letterpaper]{article}

\usepackage[pagenumbers]{cvpr} 

\usepackage{amsmath}
\usepackage{amssymb}
\usepackage{booktabs}
\usepackage{multirow}
\usepackage{lipsum}
\usepackage{algorithm}
\usepackage{algpseudocode}
\usepackage{svg}
\usepackage{overpic}
\usepackage{array}

\makeatletter
\def\thanks#1{%
  \protected@xdef\@thanks{\@thanks
    \protect\footnotetext{#1}}%
}
\makeatother

\definecolor{cvprblue}{rgb}{0.21,0.49,0.74}
\usepackage[pagebackref,breaklinks,colorlinks,allcolors=cvprblue]{hyperref}

\def\paperID{12565} 
\def\confName{CVPR}
\def\confYear{2026}

\title{RefRetouch: Personalized Image Retouching without Test-time Fine-tuning}

\author{
Temesgen Muruts Weldengus$^{1}$ \quad
Binnan Liu$^{1}$ \quad
Fei Kou$^{2}$ \quad
Youwei Lyu$^{2}$ \quad
Jinwei Chen$^{2}$ \\
Qingnan Fan$^{2\dagger}$ \quad
Changqing Zou$^{1,3\dagger}$ \\[4pt]
$^{1}$Zhejiang University \quad
$^{2}$vivo BlueImage Lab \quad
$^{3}$Zhejiang Lab \\[3pt]
{\tt\small tememuruts@zju.edu.cn} \quad
{\tt\small changqing.zou@zju.edu.cn} \quad
{\tt\small fqnchina@gmail.com} \\[3pt]
{\small
\href{https://tememuruts.github.io/RefRetouch-Project-Page/}
{\texttt{tememuruts.github.io/RefRetouch-Project-Page}}}
\thanks{$^{\dagger}$Corresponding authors.}
}

\begin{document}
\maketitle

\begin{abstract}
    Personalized image retouching aims to adapt retouching styles of individual users from reference examples, but existing methods often require user-specific fine-tuning or fail to generalize effectively. To address these challenges, we introduce \textbf{RefRetouch}, a general framework for personalized image retouching that instantly adapts to user retouching styles without any test-time fine-tuning. It employs an \textit{asymmetric auto-encoder} to encode the retouching style from paired examples into a content disentangled latent representation that enables faithful transfer of the retouching style to new images. To adaptively apply the encoded retouching style to new images, we further propose \textit{retrieval-augmented retouching} (RAR), which retrieves and aggregates style latents from reference pairs most similar in content to the query image. With these components, \textbf{RefRetouch} enables superior and generic content-aware retouching personalization across diverse scenarios, including single-reference, multi-reference, and mixed-style settings, while also generalizing out of the box to photorealistic style transfer.
\end{abstract}

\section{Introduction}
\label{sec:intro}

Image retouching enhances photographs to improve aesthetics or convey a specific narrative, but it is an inherently subjective task \cite{kim2020pienet, song2021starenhancer}, with user preferences varying widely. Existing methods \cite{gharbi2017deep, zeng2020learning, he2020conditional, yang2022adaint, ouyang2023rsfnet}, while automating the process, often learn fixed styles and require large datasets and retraining to adapt to new users, which limits their practical application. 
To address this, image retouching approaches need to be able to adapt to user preferences using only a few reference examples, enabling personalized retouching at test time without the need for retraining or extensive datasets.

\begin{figure}[htpb]
    \centering
    \includegraphics[width=\linewidth]{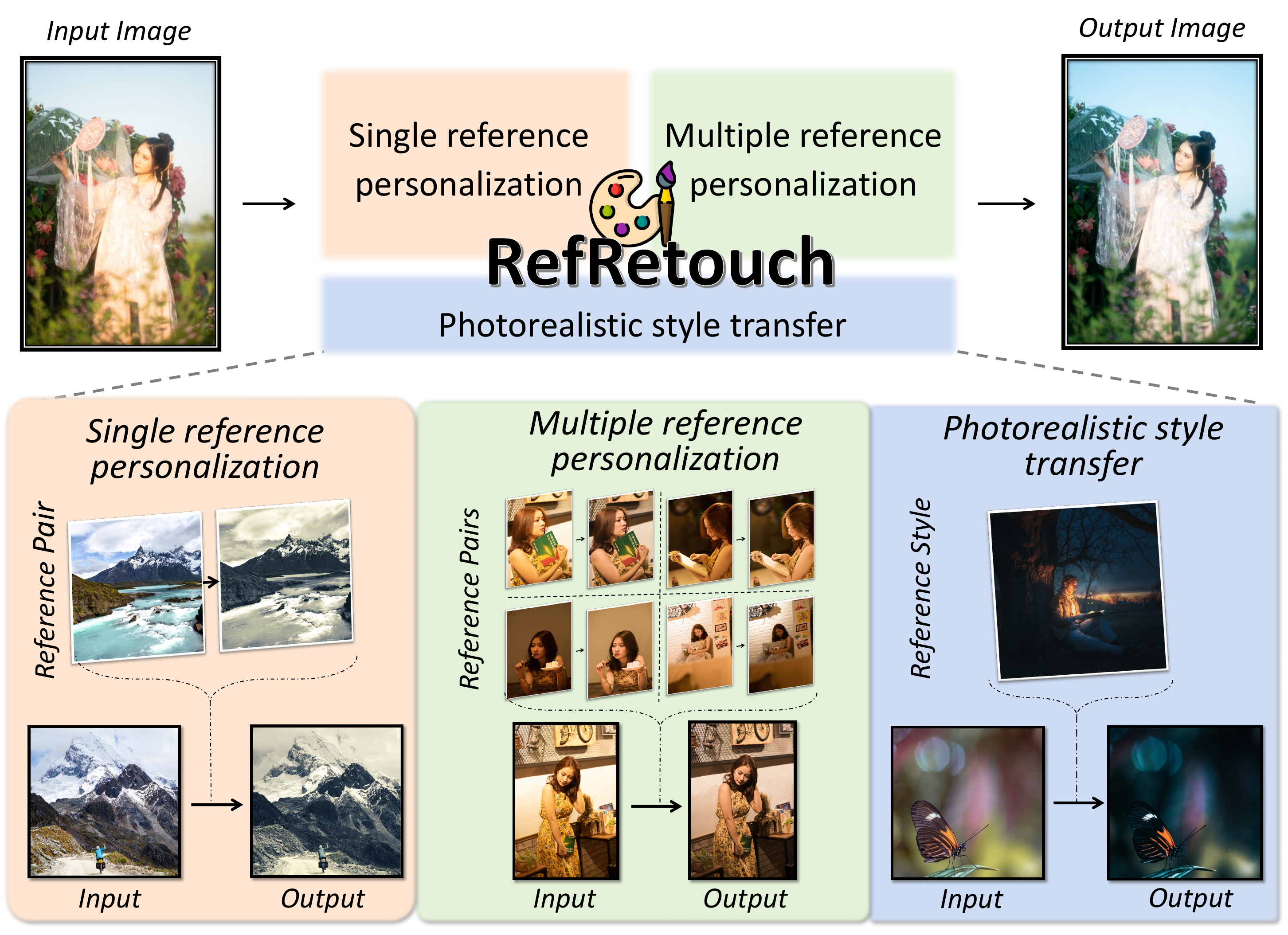}
    \caption{Our proposed method effectively personalizes image retouching from single or multiple references of similar or diverse styles at inference time, with no test-time finetuning. It also supports photorealistic style transfer out of the box.}
    \label{fig:teaser}
\end{figure}

Several works have explored generic personalized image retouching by conditioning on user-provided reference images \cite{kim2020pienet,10149499,song2021starenhancer}. While these methods demonstrate promising results, their effectiveness is fundamentally constrained by their weak representation of users' retouching styles. Specifically, the methodologies adopted by these approaches such as style classification \cite{song2021starenhancer}, metric learning \cite{kim2020pienet}, and simple auto-encoding \cite{10149499} struggle to capture retouching styles in a manner that is both disentangled from image content and transferable to unseen images. Additionally, they lack the ability to integrate information from multiple reference images, which prevents coherent retouching style fusion from multiple user references and constrains their ability to handle personalized image retouching scenarios involving diverse or mixed styles.

To overcome these limitations, we introduce \textbf{RefRetouch}, a generic personalized image retouching framework that accurately captures a user’s retouching style from a small set of reference examples and adaptively applies it to new images based on their visual content. Central to our approach is an \textit{Asymmetric Auto-Encoder} which encodes the retouching style between an original image and its retouched version into a high-level latent representation. The architecture comprises a twin encoder network with shared weights that processes the input and its corresponding retouched target, capturing the color and tone transformations while abstracting away the semantic content. This resulting latent representation is then used as a condition to control a lightweight decoder, implemented as a conditional multilayer perceptron (MLP) operating in the color domain, which faithfully reconstructs the retouched image from the original input. This design disentangles retouching style from image semantics, enabling precise and consistent transfer of the encoded retouching style to new query images without any test-time fine-tuning.  

Having encoded the retouching style of a pair of images into disentangled latent representations, we introduce \textit{retrieval-augmented retouching (RAR)} to personalize and adaptively retouch new images using style information derived from a user's preference examples. RAR performs content-aware retouching by retrieving the top-$K$ most relevant reference pairs from the user’s preference set based on visual similarity to the new input image. The corresponding retouching style latent codes of these retrieved examples are then aggregated via a weighted average, producing a composite retouch style that reflects the most relevant color and tone transformations for the current input. This aggregated retouching style latent, along with the input image, is then passed to the conditional MLP decoder to obtain the final retouched output. Through this retrieval-based conditioning, our method adapts the retouching style not only to the user's overall preferences but also to the content and context of each input image.

{RefRetouch} supports personalized image retouching using one or multiple reference examples with the same or varying styles, as shown in \Cref{fig:teaser}. Moreover, it generalizes seamlessly to photorealistic style transfer with an unpaired single reference image \cite{yoo2019photorealistic, ho2021deep, ke2023neural}, without any retraining, by simply reusing the new input image as a pseudo reference input. This demonstrates the model’s ability to encode retouching style even from semantically different images. In summary, our contributions are as follows:  

\begin{itemize}
    \item A generic framework for personalized image retouching to instantly adapt a user's retouching style from few reference examples.
    \item An Asymmetric Auto-Encoder that encodes user preferences into a content-disentangled retouching style latent representation.
    \item A retrieval-augmented retouching (RAR) module that enables content-adaptive image retouching personalization by combining relevant reference styles.
\end{itemize}

\section{Related Work}
\label{sec:related_works}

\begin{figure}
    \centering
    \includegraphics[width=\linewidth]{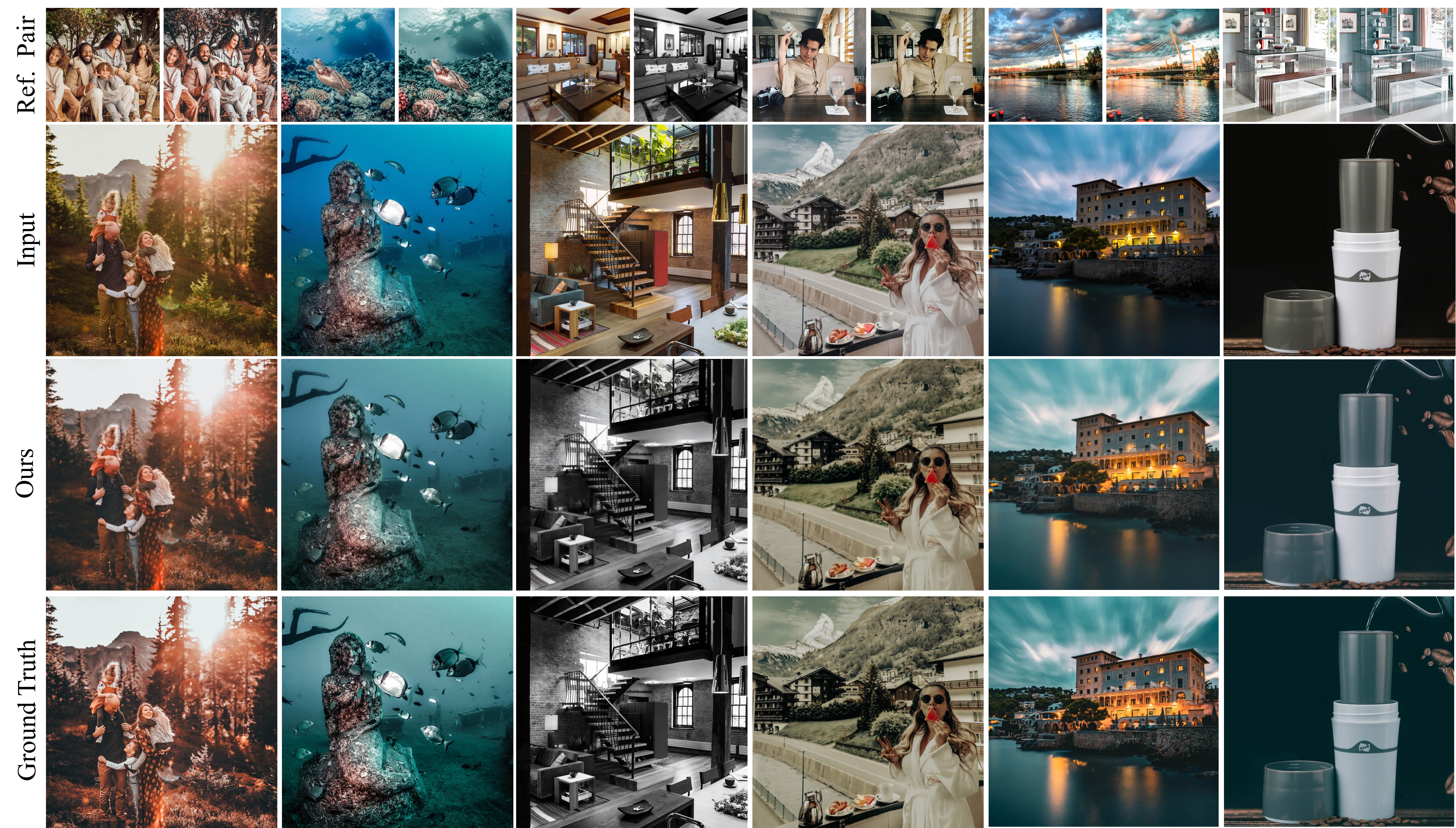}
    \caption{Image retouching style transfer from a single reference pair. \textit{Top}: input–output reference pair defining the retouching style. \textit{Second row}: query input images. \textit{Third row}: our method’s results. \textit{Bottom}: ground-truth retouched images.}
    \label{fig:single_ref_results}
\end{figure}

\textbf{Image Retouching.} Numerous approaches have been proposed to automate image retouching, ranging from traditional enhancement methods to modern learning-based techniques. Many works \cite{gharbi2017deep, zeng2020learning, yang2022adaint, moran2020deeplpf, kim2024image, serrano2024namedcurves, kim2025lightweight} optimize classical enhancement operators using paired datasets of input and expert-retouched images. Others adopt convolutional networks or multilayer perceptrons to directly map input images to retouched outputs \cite{wang2019underexposed, chen2018deep, isola2017image, wang2022neural, he2020conditional}. While effective at modeling user's retouching style, these methods apply a fixed retouching style at test time, limiting their applicability in real-world scenarios where user preferences are diverse and inherently subjective. White-box approaches \cite{hu2018exposure, ouyang2023rsfnet, shi2021learning} aim to improve user control by exposing interpretable pipelines and allowing manual adjustments to auto-retouched results. However, they still rely on models trained to mimic a specific expert's style and require active user interaction in the editing process. Methods such as \cite{elezabi2024inretouch, wu2024goal, tseng2022neural} demonstrate that fine-tuning a model on one or a few style image pairs can effectively transfer user preferences to new images. Nonetheless, these methods require re-training for each new user or style, which limits their practicality in real-world settings. In contrast to all the above methods, our approach enables test-time adaptation to user preferences without any model retraining by conditioning directly on a small set of user-provided reference examples.

\noindent\textbf{Generic personalized retouching}\quad Several methods \cite{duan2025diffretouch, wang2023learning, kim2025oneta} train a model on diverse retouching styles to produce multiple outputs from a single model at inference time. However, they do not support personalization via user-provided reference examples. Recent approaches use multi-modal large language models for text-guided image retouching \cite{chen2025photoartagent, dutt2025monetgpt, lin2025jarvisart}, but textual descriptions are often too vague and abstract to precisely capture a user's nuanced retouching preferences. Moreover, the subtle adjustments of color and tone in image retouching are difficult to describe using text. Methods like StarEnhancer \cite{song2021starenhancer}, MSM \cite{10149499}, and PieNet \cite{kim2020pienet} offer test-time personalization with a small set of user-preferred examples but they suffer from weak representations and ineffective adaptation to varying image content. Photorealistic style transfer \cite{ke2023neural, ho2021deep, yoo2019photorealistic, li2018closed} can perform single-reference color transfer but cannot handle multi-reference personalization or content-aware adaptability. Generic image editing frameworks \cite{li2025visualcloze, wang2023images, bar2022visual, chen2025edit, google2025imagegeneration, seedream2025seedream} use visual in-context learning to unify various tasks but perform poorly compared to models specialized for image retouching.

\section{Method}
\label{sec:method}

\begin{figure*}[htpb]
    \centering
    \includegraphics[width=\linewidth]{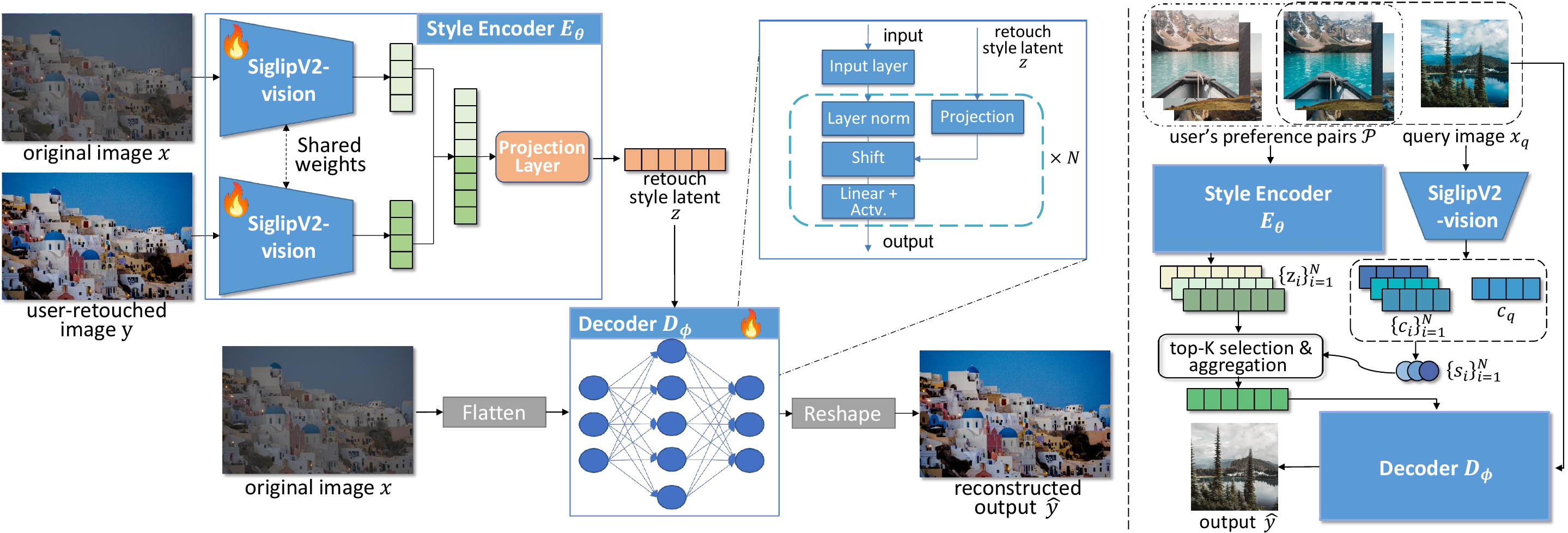}
    \caption{\textbf{Illustration of our RefRetouch pipeline.} The asymmetric auto-encoder comprises a LoRA-adapted Siamese encoder and a lightweight conditional MLP decoder. The encoder processes input-output pairs to learn a content-disentangled retouching style latent; the conditional MLP decoder then applies this latent representation to reconstruct the retouched image from the original input, operating in color space to preserve content. The retrieval-augmented retouching (RAR) module enables contextual retouching personalization by selecting and combining the most relevant latents from multiple reference images.}
    \label{fig:asymmetric_ae}
\end{figure*}

\subsection{Asymmetric Auto-Encoder}

In generic personalized image retouching, the first step is to effectively represent a user's retouching style from the reference example pairs. StarEnhancer \cite{song2021starenhancer} learns a style representation by training a classifier on a fixed set of retouching styles, but this closed-set formulation limits its ability to generalize to unseen styles. PieNet \cite{kim2020pienet} uses metric learning \cite{hoffer2015deep} to map images with similar retouching effects close in embedding space. However, the contrastive signal used in this setup is too coarse to capture the fine-grained color and lighting variation that characterizes individual user's retouching preferences. MSM \cite{10149499} adopts a U-Net with a conditional decoder to predict the residual between input and retouched images via reconstruction loss, but the resulting representation entangles style with content, preventing robust retouching style transfer to new images with different visual compositions.

To address these limitations, we propose an \textbf{asymmetric auto-encoder} whose encoder and decoder are not mirror-symmetric, both in architecture and in the spaces they operate on. The model encodes retouching style into a content-disentangled latent representation, enabling faithful reconstruction of the retouched image from its original input. As illustrated in~\Cref{fig:asymmetric_ae}, our model comprises a style encoder $E_\theta$, which extracts the retouching style from input–output image pairs, and a lightweight decoder $D_\phi$, which reconstructs the retouched image conditioned on the original input and the inferred retouching style representation. Given an original image $x \in \mathcal{X}$ and its user-retouched version $y \in \mathcal{Y}$, the encoder computes a latent vector $z = E_\theta(x, y)$, which captures high-level retouching attributes such as color and tone shifts. The decoder then synthesizes the retouched output as $\hat{y} = D_\phi(x, z)$.

The encoder is implemented as a Siamese network based on the SigLIPv2 backbone \cite{tschannen2025siglip}, processing both the original and retouched images. Features pooled from both branches are concatenated and passed through a projection layer to form the final retouching style representation $z \in \mathbb{R}^d$. To effectively adapt the encoder to retouching tasks, we add trainable LoRA weights, enabling efficient fine-tuning for retouching style representation learning.

The decoder is designed as a lightweight conditional MLP that operates in color space, mapping per-pixel values from $\mathbb{R}^3 \rightarrow \mathbb{R}^3$. This design is motivated by two key factors. First, it encourages the decoder to rely on the retouching style latent $z$ rather than the model parameters during reconstruction. It forces the latent representation to encode color and lighting transformations that are disentangled from image content, since the difference between input and retouched images is a variation in color and lighting. Second, prior works \cite{zeng2020learning, yang2022adaint, he2020conditional, conde2024nilut} have demonstrated that operating in color space yields higher image fidelity compared to spatial-domain operators, while also being resolution-agnostic and computationally efficient.

The architecture of the conditional MLP decoder is shown in~\Cref{fig:asymmetric_ae}. It consists of an input layer followed by $N$ conditional MLP blocks, inspired by feature conditioning in modern transformer \cite{peebles2023scalable}. To incorporate the retouching style latent, we project $z$ and add it to the hidden features of each MLP block. While we also experimented with more complex conditioning strategies such as adaptive layer normalization and cross-attention, the simple additive method performed best in our setting, offering a good balance between simplicity and effectiveness.

To train the encoder–decoder pair, we minimize an $\ell_1$ reconstruction loss between the predicted and ground-truth retouched images:
\begin{equation}
\mathcal{L}_{\text{recon}} = \mathbb{E}_{(x, y) \sim \mathcal{D}} \left[ \| D_\phi(x, E_\theta(x, y)) - y \|_1 \right],
\end{equation}
where $\mathcal{D}$ is a paired image retouching dataset. This objective encourages the encoder to encode retouch-specific transformations, while the decoder learns to apply these transformations faithfully to the original image. Once trained, our method can encode the user's preference pairs $\mathcal{P} = \{(x_i, y_i)\}_{i=1}^K$, into retouching style latents, precisely capturing the user's preferences. 

\subsection{Retrieval-Augmented Retouching}

After representing a user’s preferences through the retouching style latents, the next step is to apply it to a new query image. StarEnhancer \cite{song2021starenhancer} and PieNet \cite{kim2020pienet} average per-pair style embeddings to create a single user preference vector, conditioning retouching on this global code. However, this ignores the content-dependent nature of image retouching, as users apply different styles to images based on their content (e.g., overexposed vs. underexposed scenes). MSM \cite{10149499} models content-dependent retouching using a transformer that predicts the query style from the reference content/style embeddings and the query content embedding, and is trained through a meta-learning framework. However, this framework relies on unpaired user images, where pseudo input--output pairs are constructed by synthetically degrading the unpaired images. As a result, the target outputs often exhibit relatively homogeneous color and tone distributions, while variation is primarily introduced only on the input side. This constrains the learned mapping space, encouraging the model to reproduce the output distribution of the training images rather than learning a robust content-aware style prediction capability.

To address these limitations, we propose \textbf{retrieval-augmented retouching (RAR)}, which applies retouching styles by retrieving reference examples that are most relevant to a given query image, as illustrated on the right side of~\Cref{fig:asymmetric_ae}. Given a query image $x_{\text{q}}$, we first extract its content embedding $c_{\text{q}}$, together with the content embeddings $\{c_i\}$ of the input images in the reference. Specifically, we use the LoRA-adapted SigLIPv2 backbone from the style encoder and take its pooled features, which are shown to capture retouching-relevant content: color and tone cues, as detailed in~\Cref{sec:photometric_encoder}. For each reference pair, we also encode its retouching style latent $z_i$ using the style encoder, representing the user's retouching preference. We then compute the relevance between the query image and each reference input by measuring the cosine similarity between their content embeddings: $s_i = \frac{\mathbf{c}_{\text{q}} \cdot \mathbf{c}_i}{\|\mathbf{c}_{\text{q}}\| \|\mathbf{c}_i\|}.$ Based on these similarity scores, we retrieve the top-$K$ nearest reference examples, denoted as $\mathcal{N}_K$, and aggregate their corresponding retouching style latents through a softmax-weighted combination:

\begin{equation}\label{eq:rar}
    w_i=\frac{\exp\!\left(\frac{s_i}{\tau}\right)}{\sum_{j\in \mathcal{N}_K}\exp\!\left(\frac{s_j}{\tau}\right)}, 
\qquad 
z_{\text{q}}=\sum_{i\in \mathcal{N}_K} w_i\, z_i,
\end{equation}

\noindent where $\tau>0$ is the temperature controlling the sharpness of the weighting. The personalized retouching result is then obtained by the conditional decoder as,
\begin{equation}
\hat{y}_{\text{q}} = D_{\phi}(x_{\text{q}}, z_{\text{q}}),
\end{equation}
applying the aggregated style $z_{\text{q}}$ to the query image. By retrieving and weighting styles based on content similarity, RAR adapts user preferences to each image’s characteristics rather than enforcing a single global style for every input image.

\section{Experiments}
\label{sec:experiments}
\providecommand{\rotup}[1]{\rotatebox[origin=c]{90}{#1}}

\subsection{Dataset}
To train the asymmetric auto-encoder in our \textbf{RefRetouch}, we collect 800 free presets from the Adobe Lightroom community \cite{adobe_lightroom} and apply them to 95,000 images sampled from the LAION dataset \cite{schuhmann2022laion}. This process provides a diverse set of real-world retouching transformations while maintaining a clear separation between training and evaluation domains. To ensure fair benchmarking, we exclude all evaluation benchmark images and presets from the training set, in contrast to previous methods that incorporated benchmark content or styles during training \cite{song2021starenhancer, kim2020pienet}. Additional details on the dataset construction, filtering, and preset selection are provided in the supplementary material.

\subsection{Experimental Setup}  
We use SigLIPv2 as the backbone of the Siamese style encoder and add LoRA adapters (rank 16) to all linear layers in the attention blocks. The projection layer takes the concatenation of the features pooled from the two branches and outputs a 2048-dimensional retouching style latent. The conditional MLP decoder operates per pixel in color space. It consists of an input layer that maps an RGB pixel to a 128-dimensional hidden feature, followed by three conditional MLP blocks with hidden dimensions $[256,\,512,\,3]$, as shown in~\Cref{fig:asymmetric_ae}. All intermediate layers use ReLU activations, while the final layer uses a sigmoid to constrain outputs to $[0,1]$. The retouching style latent is injected into all hidden layers of the MLP decoder after projecting it using a Linear+ReLU to match the target layer's hidden dimension. This model is then trained for $180,000$ steps with a batch size of 8 on $384\times384$ random crops of the training images. For content embeddings in retrieval-augmented retouching (RAR), we reuse the LoRA-adapted SigLIPv2 backbone and extract its pooled features. Then, we compute cosine similarity between the query and references' input embeddings to retrieve the most relevant references using~\Cref{eq:rar}. Unless stated otherwise, we set the softmax temperature to $\tau = 0.1$, and $K=3$ in all the experiments.

\subsection{Benchmarks}
We evaluate image retouching personalization across three increasingly challenging settings: (1) single-style retouching, (2) multi-style retouching with consistent edits, and (3) multi-style retouching with inconsistent, implicitly defined edits.

\noindent \textbf{VCIRB (Single-Style Retouching)}\quad To evaluate personalization under a single, consistent retouching style, we construct the Visually Consistent Image Retouching Benchmark (VCIRB) by applying 25 unseen Adobe Lightroom presets to images sampled from LAION \cite{schuhmann2022laion}. Since a preset may produce inconsistent results across diverse content \cite{elezabi2024inretouch, ke2023neural}, we cluster images into 25 semantically and color-tone coherent groups using SigLIPv2 embeddings and $ab$-channel histograms in CIELAB space. Presets are applied within these clusters to ensure visual consistency in the retouched outputs. We compare against several generic personalized image retouching baselines: StarEnhancer \cite{song2021starenhancer}, MSM~\cite{10149499}, VisualCloze \cite{li2025visualcloze} (a generic in-context generative editing model), PhotoArtAgent \cite{chen2025photoartagent}, Seedream4.0~\cite{seedream2025seedream}, and NanoBanana \cite{google2025imagegeneration}. PhotoArtAgent, Seedream 4.0, and NanoBanana are evaluated only on single-pair personalization, as they are constrained by a short effective context window, where using more than two images as input leads to degraded performance. We also include retrained variants of MSM and VisualCloze, trained on our meta-learning dataset in which 800 Lightroom presets are applied to 95,000 LAION images clustered by semantic content and tonal similarity, following the same protocol as VCIRB (see supplementary material for dataset construction details).

\noindent \textbf{PPR10K-Groups (Multi-Style, Consistent Retouching)}\quad In real-world scenarios such as event or portrait photography, users often retouch related images into a consistent output style. The PPR10K dataset \cite{liang2021ppr10k} groups portrait images of the same subjects captured in different scenes, along with their expert-retouched versions, where experts are instructed to apply consistent adjustments across all images within a group. This inherent style consistency allows a few reference samples from a group to effectively capture the user’s retouching preferences, while the variation in scene content provides a natural testbed for evaluating a model’s context-awareness. We select 21 PPR10K groups, retouched by expert A, with more than 12 images each and sample $1$, $3$, or $6$ reference pairs per group and test on the remaining images. We compare with the same baselines as in VCIRB, testing each model’s ability to handle multiple references in a multi-style setting.

\noindent \textbf{MIT-FiveK and PPR10K Users (Multi-Style, Inconsistent Retouching)}\quad Another common use case is adapting to a user's retouching style from historical edits, which is often diverse and inconsistent. This setting is widely used in prior personalization works \cite{10149499, yang2022adaint, ouyang2023rsfnet}. We follow the standard protocol on MIT-Adobe FiveK \cite{fivek} and PPR10K \cite{liang2021ppr10k} to split the datasets into training and test splits. From each training split, we sample $20$, $50$, and $100$ input–output reference pairs and evaluate on the test sets ($498$ for FiveK and $2{,}286$ for PPR10K). Unlike prior work \cite{10149499}, we explicitly sample references based on color and tone diversity to minimize distribution mismatch between reference and test images. These fixed references are then used for all methods. Further details of the sampling strategy are provided in the supplementary material. To compare our method, we select baselines capable of handling more than 20 references. While VisualCloze can theoretically accommodate any number of references, its computational complexity increases rapidly as the number of references grows, since it directly processes the patchified 2D latent encodings of the reference pairs using a large diffusion transformer. Therefore, we exclude it from comparison in this setting. Additionally, we include two state-of-the-art, training-based image personalization methods: AdaInt \cite{yang2022adaint} and RSFNet \cite{ouyang2023rsfnet} to assess and compare their performance when training data is limited. Both are trained on the reference images for 100 epochs using their default configurations. Note that we exclude StarEnhancer from the evaluation on the MIT-FiveK data because the model was already trained on MIT-FiveK.

All images in VCIRB are resized to $512\times512$, while the original resolution is maintained for images from MIT-FiveK and PPR10K. For VisualCloze, we pre-resize the images to be divisible by 16 using a right crop, and evaluate them at this resolution to prevent loss due to VAE downsampling and patchification. PSNR, SSIM, and LPIPS are used to evaluate image retouching personalization performance across all three benchmarks. The reported metrics are averaged over all test images and across groups (if applicable).


\begin{figure*}[htpb]
    \centering
    \scriptsize
    \setlength\tabcolsep{0.1pt}
\begin{tabular}{cccccc} %
    

    Input & StarEnhancer~\cite{song2021starenhancer} & MSM~\cite{10149499} & VisualCloze~\cite{li2025visualcloze} & Ours & Target \\ 
    
    \begin{subfigure}[c]{0.166\textwidth}
        \centering
        \begin{overpic}[percent,width=\textwidth]{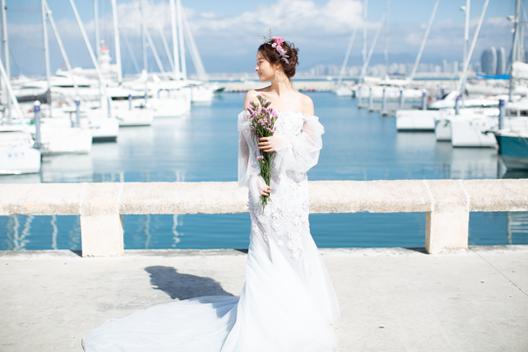}
        \end{overpic}
    \end{subfigure}  
    &
    \begin{subfigure}[c]{0.166\textwidth}
        \centering
        \begin{overpic}[percent,width=\textwidth]{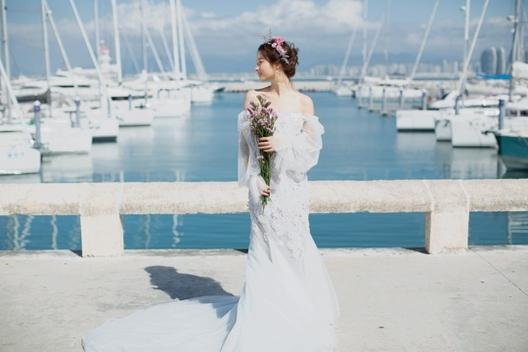}
        \end{overpic}
    \end{subfigure}  
    &
    \begin{subfigure}[c]{0.166\textwidth}
        \centering
        \begin{overpic}[percent,width=\textwidth]{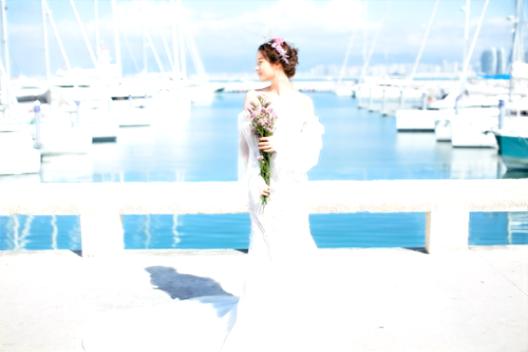}
        \end{overpic}
    \end{subfigure}  
    &
    \begin{subfigure}[c]{0.166\textwidth}
        \centering
        \begin{overpic}[percent,width=\textwidth]{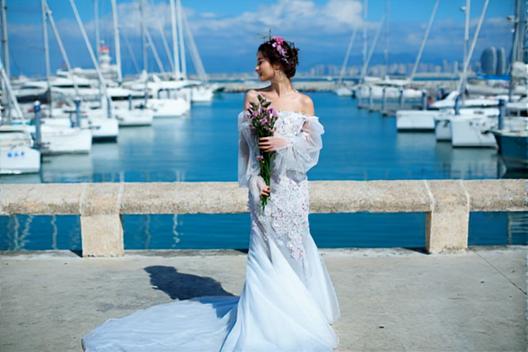}
        \end{overpic}
    \end{subfigure}  
    &
    \begin{subfigure}[c]{0.166\textwidth}
        \centering
        \begin{overpic}[percent,width=\textwidth]{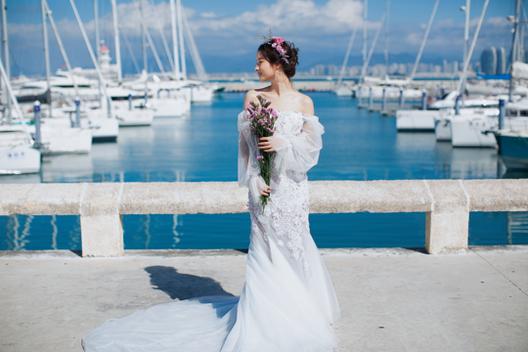}
        \end{overpic}
    \end{subfigure}  
    &
    \begin{subfigure}[c]{0.166\textwidth}
        \centering
        \begin{overpic}[percent,width=\textwidth]{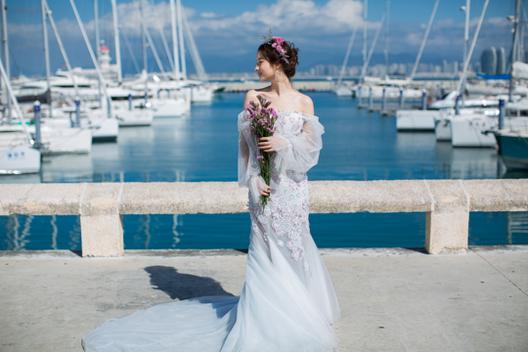}
        \end{overpic}
    \end{subfigure}  
    \\
    PSNR/SSIM & 18.23/0.906  & 
    9.33/0.667 & 23.88/0.924 & 27.44/0.987 & 
    \\
    \begin{subfigure}[c]{0.166\textwidth}
        \centering
        \begin{overpic}[percent,width=\textwidth]{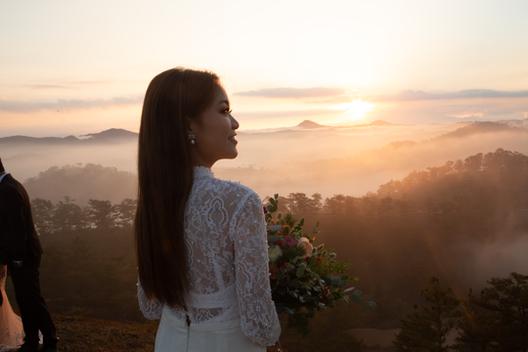}
        \end{overpic}
    \end{subfigure}  
    &
    \begin{subfigure}[c]{0.166\textwidth}
        \centering
        \begin{overpic}[percent,width=\textwidth]{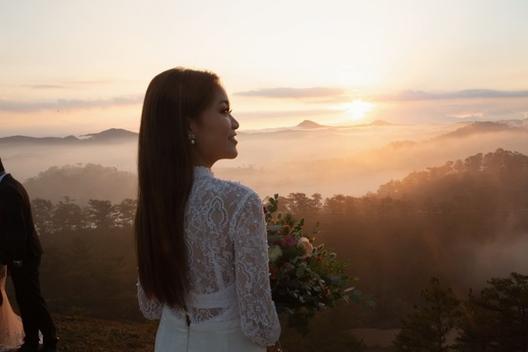}
        \end{overpic}
    \end{subfigure}  
    &
    \begin{subfigure}[c]{0.166\textwidth}
        \centering
        \begin{overpic}[percent,width=\textwidth]{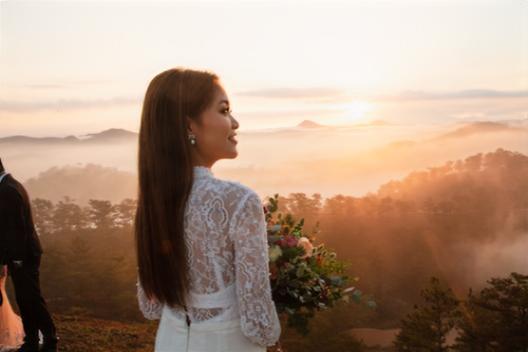}
        \end{overpic}
    \end{subfigure}  
    &
    \begin{subfigure}[c]{0.166\textwidth}
        \centering
        \begin{overpic}[percent,width=\textwidth]{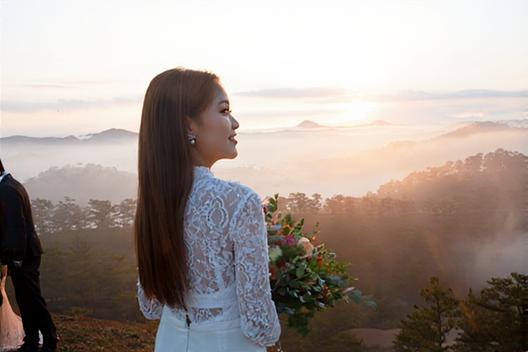}
        \end{overpic}
    \end{subfigure}  
    &
    \begin{subfigure}[c]{0.166\textwidth}
        \centering
        \begin{overpic}[percent,width=\textwidth]{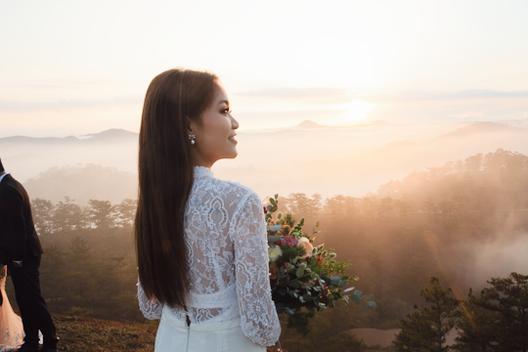}
        \end{overpic}
    \end{subfigure}  
    &
    \begin{subfigure}[c]{0.166\textwidth}
        \centering
        \begin{overpic}[percent,width=\textwidth]{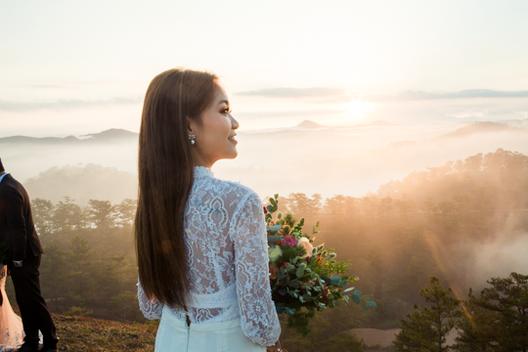}
        \end{overpic}
    \end{subfigure}  
    \\
    PSNR/SSIM & 14.20/0.81  & 
    20.17/0.938 & 22.08/0.917 & 30.81/0.974 & 
     \\
    \begin{subfigure}[c]{0.166\textwidth}
        \centering
        \begin{overpic}[percent,width=\textwidth]{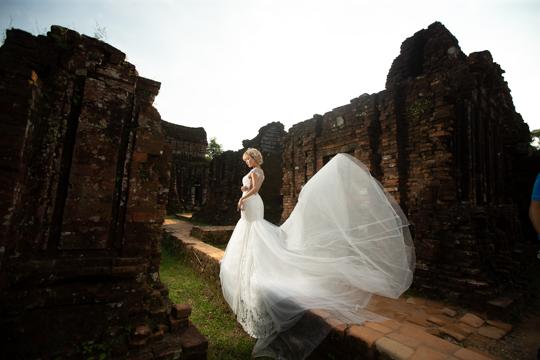}
        \end{overpic}
    \end{subfigure}  
    &
    \begin{subfigure}[c]{0.166\textwidth}
        \centering
        \begin{overpic}[percent,width=\textwidth]{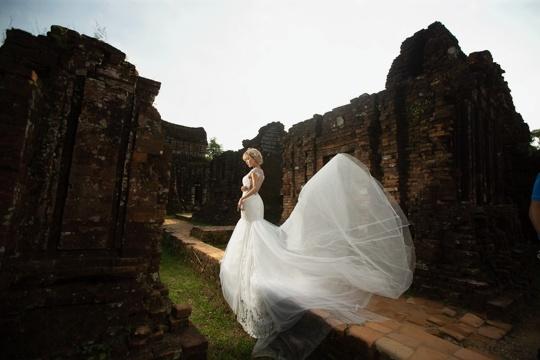}
        \end{overpic}
    \end{subfigure}  
    &
    \begin{subfigure}[c]{0.166\textwidth}
        \centering
        \begin{overpic}[percent,width=\textwidth]{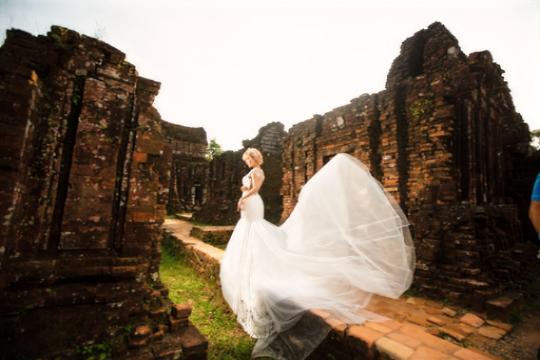}
        \end{overpic}
    \end{subfigure}  
    &
    \begin{subfigure}[c]{0.166\textwidth}
        \centering
        \begin{overpic}[percent,width=\textwidth]{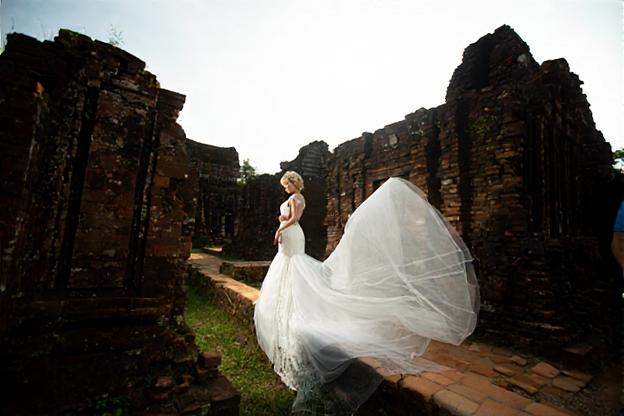}
        \end{overpic}
    \end{subfigure}  
    &
    \begin{subfigure}[c]{0.166\textwidth}
        \centering
        \begin{overpic}[percent,width=\textwidth]{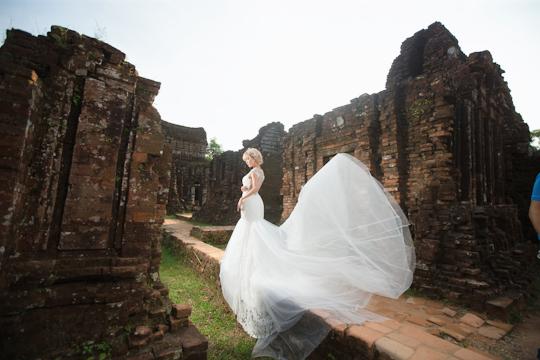}
        \end{overpic}
    \end{subfigure}  
    &
    \begin{subfigure}[c]{0.166\textwidth}
        \centering
        \begin{overpic}[percent,width=\textwidth]{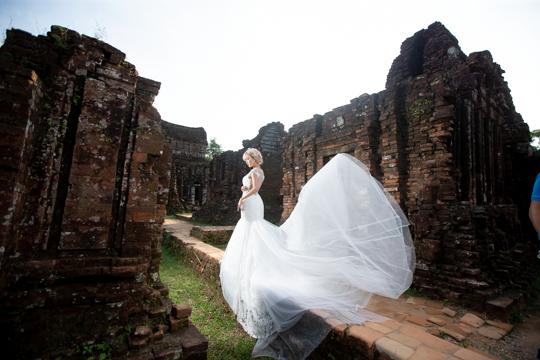}
        \end{overpic}
    \end{subfigure}  
    \\
    PSNR/SSIM & 19.84/0.833 & 25.42/0.929 & 20.27/0.746 & 27.52/0.899  & 
    \\
    \begin{subfigure}[c]{0.166\textwidth}
        \centering
        \begin{overpic}[percent,width=\textwidth]{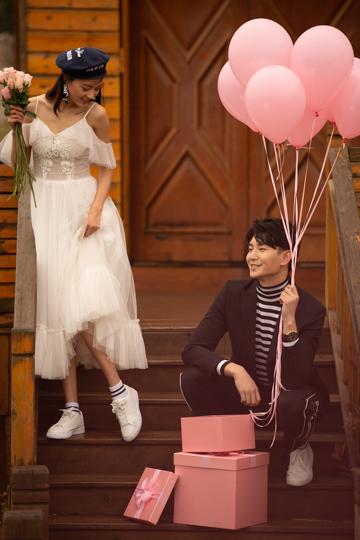}
        \end{overpic}
    \end{subfigure}  
    &
    \begin{subfigure}[c]{0.166\textwidth}
        \centering
        \begin{overpic}[percent,width=\textwidth]{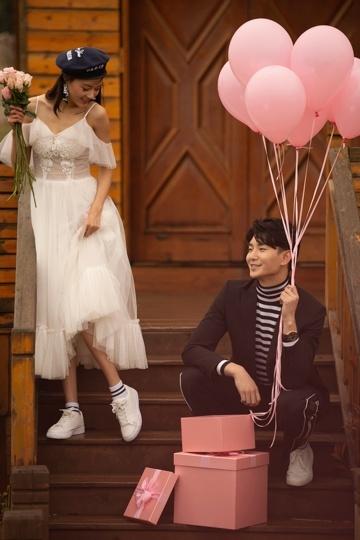}
        \end{overpic}
    \end{subfigure}  
    &
    \begin{subfigure}[c]{0.166\textwidth}
        \centering
        \begin{overpic}[percent,width=\textwidth]{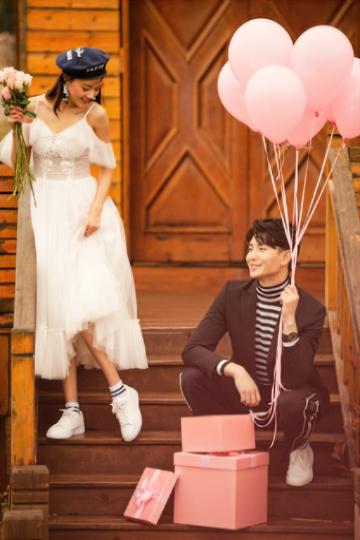}
        \end{overpic}
    \end{subfigure}  
    &
    \begin{subfigure}[c]{0.166\textwidth}
        \centering
        \begin{overpic}[percent,width=\textwidth]{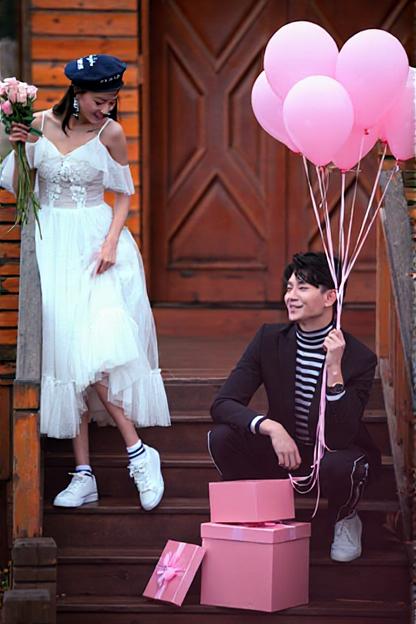}
        \end{overpic}
    \end{subfigure}  
    &
    \begin{subfigure}[c]{0.166\textwidth}
        \centering
        \begin{overpic}[percent,width=\textwidth]{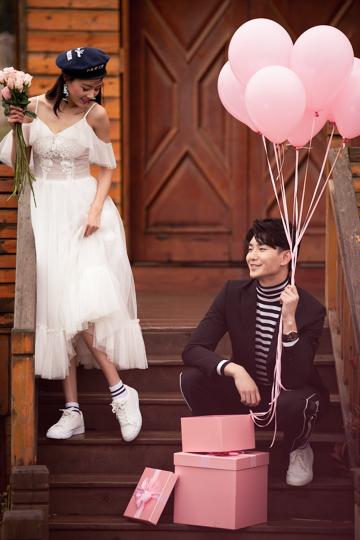}
        \end{overpic}
    \end{subfigure}  
    &
    \begin{subfigure}[c]{0.166\textwidth}
        \centering
        \begin{overpic}[percent,width=\textwidth]{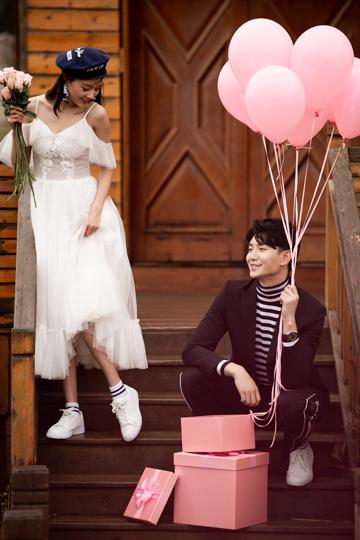}
        \end{overpic}
    \end{subfigure}  
    \\
    PSNR/SSIM & 22.38/0.946  & 18.83/0.883 & 24.68/0.946 & 32.51/0.985 & 
     \\
    \begin{subfigure}[c]{0.166\textwidth}
        \centering
        \begin{overpic}[percent,width=\textwidth]{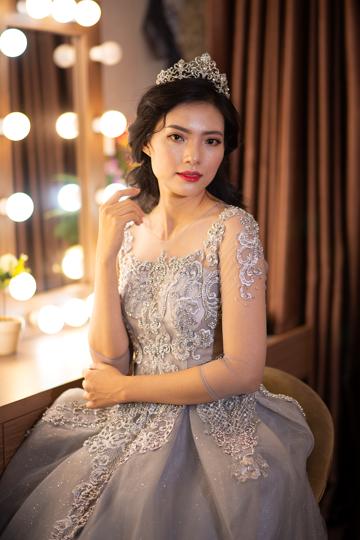}
        \end{overpic}
    \end{subfigure}  
    &
    \begin{subfigure}[c]{0.166\textwidth}
        \centering
        \begin{overpic}[percent,width=\textwidth]{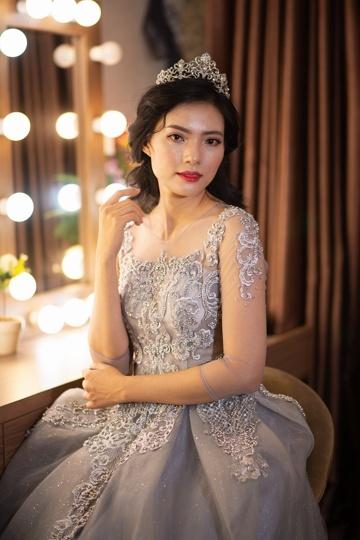}
        \end{overpic}
    \end{subfigure}  
    &
    \begin{subfigure}[c]{0.166\textwidth}
        \centering
        \begin{overpic}[percent,width=\textwidth]{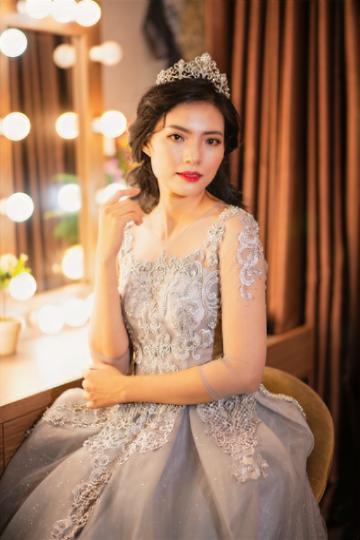}
        \end{overpic}
    \end{subfigure}  
    &
    \begin{subfigure}[c]{0.166\textwidth}
        \centering
        \begin{overpic}[percent,width=\textwidth]{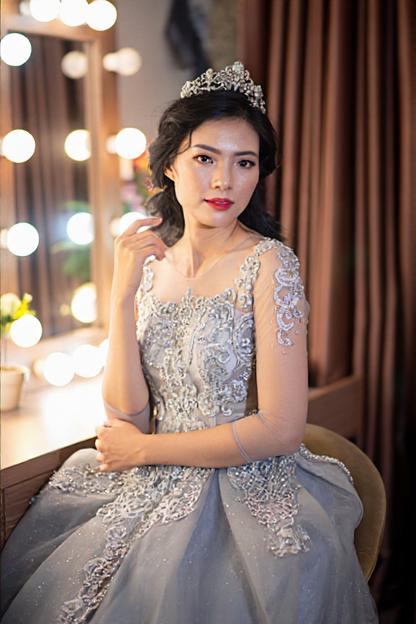}
        \end{overpic}
    \end{subfigure}  
    &
    \begin{subfigure}[c]{0.166\textwidth}
        \centering
        \begin{overpic}[percent,width=\textwidth]{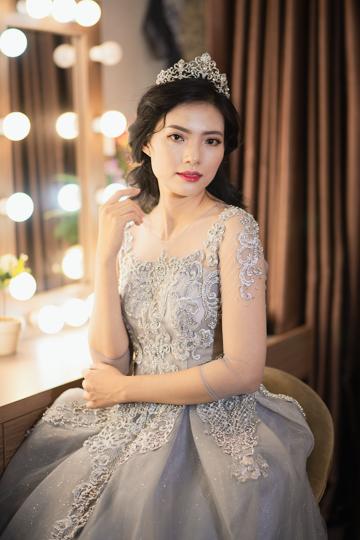}
        \end{overpic}
    \end{subfigure}  
    &
    \begin{subfigure}[c]{0.166\textwidth}
        \centering
        \begin{overpic}[percent,width=\textwidth]{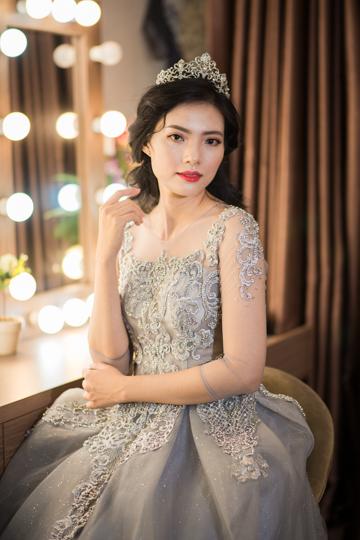}
        \end{overpic}
    \end{subfigure}  
    \\
    PSNR/SSIM & 24.30/0.962  & 24.44/0.955 & 26.48/0.924 & 32.70/0.988 & 
    
\end{tabular}

    \caption{PPR10K groups performance comparison when three references are used.}
    \label{fig:ppr10k_gropus}
\end{figure*}

\subsection{Results}

\begin{table}[htpb]
  \centering
  \caption{Reconstruction fidelity comparison. \textsuperscript{\dag} indicates that the baseline model is trained on our dataset.}
  \small
  \begin{tabular}{llccc}
    \toprule
    \textbf{Dataset} & \textbf{Method} & \textbf{PSNR$\uparrow$} & \textbf{SSIM$\uparrow$} & \textbf{LPIPS$\downarrow$} \\
    \midrule
    \multirow{3}{*}{VCIRB} & MSM \cite{10149499} & 27.28 & 0.900 & 0.099 \\
    & MSM\textsuperscript{\dag} \cite{10149499} & 29.61 & 0.923 & 0.069 \\
    & Ours & \textbf{32.42} & \textbf{0.974} & \textbf{0.035} \\
    \midrule
    \multirow{3}{*}{MIT-FiveK} & MSM \cite{10149499} & 28.94 & 0.851 & 0.137 \\
    & MSM\textsuperscript{\dag} \cite{10149499} & 26.32 & 0.803 & 0.199 \\
    & Ours & \textbf{32.95} & \textbf{0.972} & \textbf{0.028} \\
    \midrule
    \multirow{3}{*}{PPR10K-A} & MSM \cite{10149499} & 29.28 & 0.943 & 0.075 \\
    & MSM\textsuperscript{\dag} \cite{10149499} & 29.71 & 0.936 & 0.086 \\
    & Ours & \textbf{34.14} & \textbf{0.984} & \textbf{0.019} \\
    \bottomrule
  \end{tabular}
  \label{tab:recon_fidelity}
\end{table}

\begin{table}[htpb]
  \centering
  \caption{Photorealistic style transfer quantitative results. Metrics follow \cite{ke2023neural}. Higher Content and Style Scores indicate better structure preservation and style fidelity, respectively.}
  \small
  \begin{tabular}{lcc}
    \toprule
    \textbf{Method} & \textbf{Content Score}$\uparrow$ & \textbf{Style Score}$\uparrow$ \\
    \midrule
    NeuralPreset \cite{ke2023neural} & 0.856 & 0.555 \\
    DeepPreset \cite{ho2021deep} & 0.913 & 0.616 \\
    D-LUT \cite{li2025d} & 0.787 & 0.657 \\
    PhotoArtAgent \cite{chen2025photoartagent} & 0.860 & 0.530 \\
    PhotoWCT2 \cite{chiu2022photowct2} & 0.749 & 0.916 \\
    NanoBanana \cite{google2025imagegeneration} & 0.658 & 0.610 \\
    Seedream4.0 \cite{seedream2025seedream} & 0.454 & 0.461 \\
    Ours & 0.742 & 0.917 \\
    \bottomrule
  \end{tabular}
  \label{tab:target_only_scores}
\end{table}

\begin{table*}[htpb]
  \centering
  \caption{Image retouching personalization comparison on the VCIRB and PPR10K groups benchmarks. The best results are shown in bold. \textsuperscript{\dag}~indicates methods that are re-trained (or fine-tuned) on our dataset. }
  \small
  \begin{tabular}{llccccccccc}
    \toprule
    \multirow{2}{*}{\textbf{Dataset}} & \multirow{2}{*}{\textbf{Method}} & \multicolumn{3}{c}{\textbf{PSNR$\uparrow$}} & \multicolumn{3}{c}{\textbf{SSIM$\uparrow$}} & \multicolumn{3}{c}{\textbf{LPIPS$\downarrow$}} \\
    \cmidrule(lr){3-5} \cmidrule(lr){6-8} \cmidrule(lr){9-11}
    & & \textbf{1} & \textbf{2} & \textbf{4} & \textbf{1} & \textbf{2} & \textbf{4} & \textbf{1} & \textbf{2} & \textbf{4} \\
    \midrule
    \multirow{9}{*}{VCIRB} & StarEnhancer \cite{song2021starenhancer} & 21.00 & 20.98 & 21.27 & 0.864 & 0.863 & 0.868 & 0.123 & 0.123 & 0.120 \\
    & MSM \cite{10149499} & 17.69 & 18.17 & 18.53 & 0.822 & 0.830 & 0.836 & 0.164 & 0.155 & 0.150 \\
    & MSM\textsuperscript{\dag} \cite{10149499} & 24.35 & 25.03 & 25.28 & 0.900 & 0.904 & 0.905 & 0.097 & 0.094 & 0.093 \\
    & VisualCloze \cite{li2025visualcloze} & 19.81 & 18.60 & 12.05 & 0.765 & 0.731 & 0.378 & 0.173 & 0.191 & 0.591 \\ 
    & VisualCloze\textsuperscript{\dag} \cite{li2025visualcloze} & 23.91 & 24.63 & 24.70 & 0.83 & 0.834 & 0.825 & 0.077 & 0.072 & 0.079 \\ 
    & PhotoArtAgent \cite{chen2025photoartagent} & 20.03 & - & -  & 0.83 & - & - &  0.158 & - & - \\ 
    & NanoBanana \cite{google2025imagegeneration}& 11.61 & - & - & 0.387 & - & - & 0.570 & - & - \\ 
    & Seedream4.0 \cite{seedream2025seedream} & 11.8 & - & - & 0.422 & - & - & 0.488 & - & - \\ 
    & Ours & \textbf{29.13} & \textbf{29.57} & \textbf{29.82} & \textbf{0.963} & \textbf{0.964} & \textbf{0.965} & \textbf{0.048} & \textbf{0.047} & \textbf{0.046} \\
    \midrule
    & & \textbf{1} & \textbf{3} & \textbf{6} & \textbf{1} & \textbf{3} & \textbf{6} & \textbf{1} & \textbf{3} & \textbf{6} \\
    \cmidrule(lr){3-5} \cmidrule(lr){6-8} \cmidrule(lr){9-11}
    \multirow{9}{*}{PPR10K-groups} & StarEnhancer \cite{song2021starenhancer} & 19.16 & 18.83 & 19.36 & 0.875 & 0.857 & 0.866 & 0.102 & 0.108 & 0.119 \\
    & MSM \cite{10149499} & 18.86 & 18.82 & 18.59 & 0.877 & 0.873 & 0.878 & 0.154 & 0.159 & 0.155 \\
    & MSM\textsuperscript{\dag} \cite{10149499} & 19.27 & 19.62 & 18.16 & 0.876 & 0.880 & 0.866 & 0.149 & 0.148 & 0.170 \\
    & VisualCloze \cite{li2025visualcloze} & 15.04 & 14.21 & 7.74 & 0.453 & 0.432 & 0.246 & 0.22  & 0.261 & 0.844 \\ 
    & VisualCloze\textsuperscript{\dag} \cite{li2025visualcloze} & 21.12 & 21.42 & 20.98 & 0.832 & 0.832 & 0.812 & 0.097 & 0.093 & 0.112 \\ 
    & PhotoArtAgent \cite{chen2025photoartagent} & 20.01 & - & - & 0.888 & - & - & 0.093 & - & - \\
    & NanoBanana \cite{google2025imagegeneration} & 12.89 & - & - & 0.48 & - & - & 0.434 & - & - \\ 
    & Seedream4.0 \cite{seedream2025seedream} & 9.34 & - & - & 0.278 & - & - & 0.607 & - & - \\ 
   & Ours & \textbf{22.51} & \textbf{23.92} & \textbf{25.27} & \textbf{0.932} & \textbf{0.949} & \textbf{0.961} & \textbf{0.065} & \textbf{0.052} & \textbf{0.043} \\
    \bottomrule
  \end{tabular}
  \label{tab:single_style_quant}
\end{table*}

\begin{figure*}[htpb]
    \centering
    \includegraphics[width=0.98\textwidth]{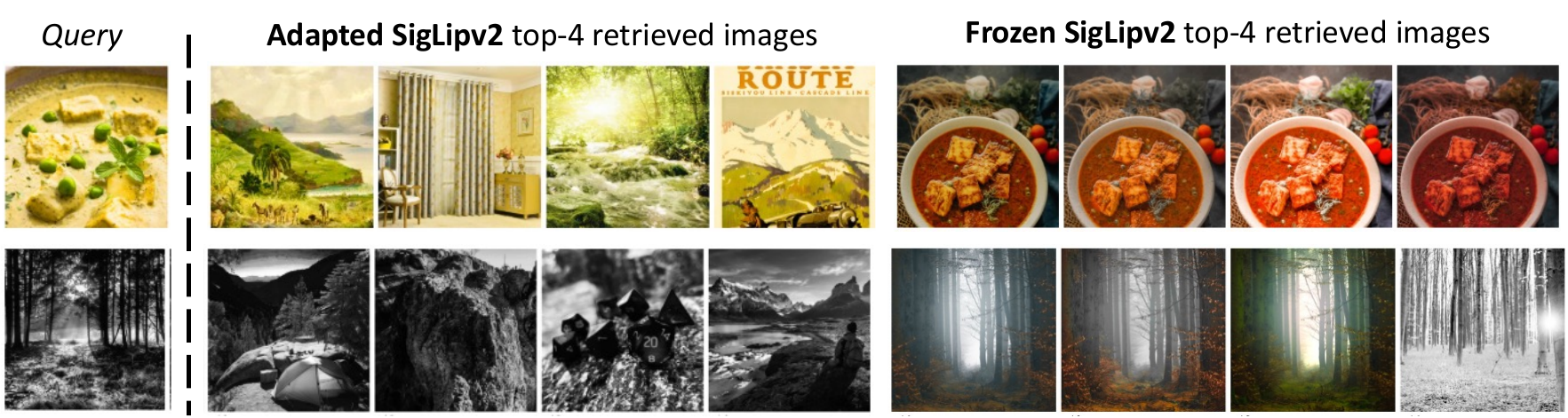}
    \caption{Feature embedding analysis through top-$K$ retrieval.}
    \label{fig:retrieval}
\end{figure*}

\textbf{Reconstruction}\quad As shown in~\Cref{tab:recon_fidelity}, we compare the reconstruction fidelity of RefRetouch with MSM \cite{10149499}, an approach that learns a retouching style latent through a reconstruction objective. We evaluate both the original MSM checkpoint and another version trained on the same dataset as ours for a fair comparison. Across the VCIRB benchmark and paired examples from MIT-Adobe FiveK \cite{fivek} and PPR10K \cite{liang2021ppr10k}, RefRetouch consistently outperforms both MSM variants. These results demonstrate that our model (i) encodes retouching styles effectively into a low-dimensional latent vector and (ii) faithfully reconstructs the retouched output from the original image and the learned latent representation.

\noindent \textbf{Single-reference retouching personalization}\quad Reconstruction alone is not sufficient; the learned retouching style latent must be disentangled from image content to enable style transfer to new query images, a key requirement for personalized retouching. To evaluate this, we extract the retouching style latent from a single reference pair and apply it to unseen query images. As shown in~\Cref{fig:single_ref_results}, RefRetouch accurately transfers the retouching style from the reference to the query images using only the latent representation. These results indicate that the encoded latent representation is effectively disentangled from image content, enabling it to capture the retouching style in the reference pair while remaining invariant to scene semantics. Quantitative results on the VCIRB benchmark in~\Cref{tab:single_style_quant} further support this finding; our method significantly outperforms existing generic retouching transfer approaches. For evaluations with multiple reference pairs, we use retrieval-augmented retouching to aggregate style cues from relevant references. However, increasing the number of references results in only marginal improvements on VCIRB, which is expected since all images in a given VCIRB group are retouched with the same preset, making a single reference sufficient to capture the style.

\noindent \textbf{Multi-reference personalization with consistent retouching styles}\quad  \Cref{tab:single_style_quant} provides the performance on the PPR10K-groups dataset, and similarly, our method consistently outperforms existing generic retouching personalization approaches. Notably, performance improves as more reference examples are added, in contrast to the high inconsistency seen in other methods. This highlights the effectiveness of our retrieval-augmented retouching in leveraging multiple references for content-aware retouching, adapting the style based on the input image's context. Qualitative comparisons using three reference examples are presented in~\Cref{fig:ppr10k_gropus}.

\noindent \textbf{Personalization with diverse and inconsistent retouching styles}\quad  The results in~\Cref{tab:standard_bench_results} show the performance of personalizing to MIT-FiveK and PPR10K users' retouching style. Our approach surpasses all generic personalized retouching baselines and achieves performance comparable to, or even exceeding, methods explicitly trained on the specific individual user references \cite{yang2022adaint, ouyang2023rsfnet}, despite being entirely tuning-free. This highlights the strong personalization capability of our approach and its ability to generalize to challenging image retouching personalization scenarios. Consistent with the PPR10K-groups evaluation, performance improves with additional reference images, highlighting the model’s ability to leverage multiple examples, especially in cases involving diverse retouching styles.

\begin{figure}[htpb]
    \centering
    \resizebox{\columnwidth}{!}{

\newcommand{\figlabel}[1]{%
\mbox{\scalebox{0.82}{\scriptsize #1}}%
}

\setlength{\tabcolsep}{1.3pt}

\begin{tabular}{@{}*{5}{>{\centering\arraybackslash}m{0.17\columnwidth}}@{}}

\figlabel{Content Image} &
\figlabel{Style Image} &
\figlabel{NeuralPreset~\cite{ke2023neural}} &
\figlabel{DeepPreset~\cite{ho2021deep}} &
\figlabel{D-LUT~\cite{li2025d}} \\[1pt]

\includegraphics[width=\linewidth]{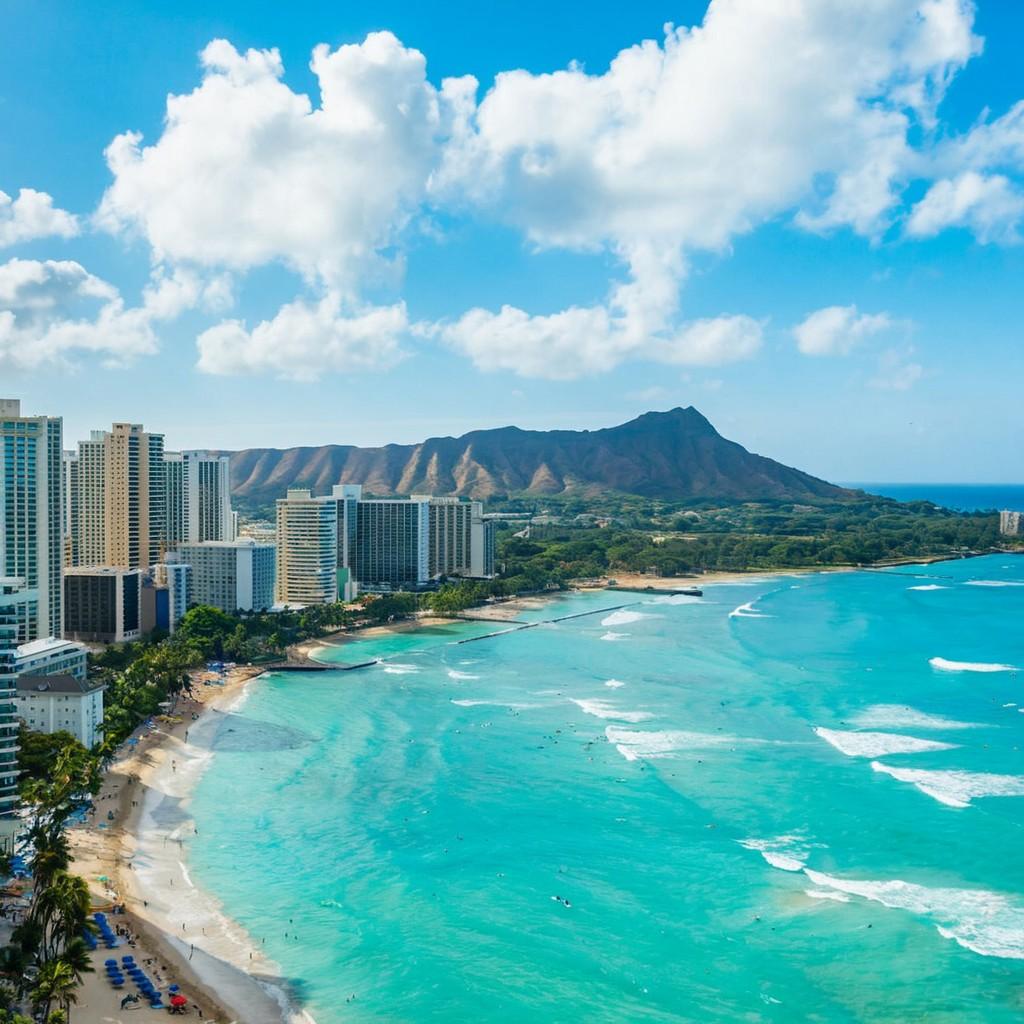} &
\includegraphics[width=\linewidth]{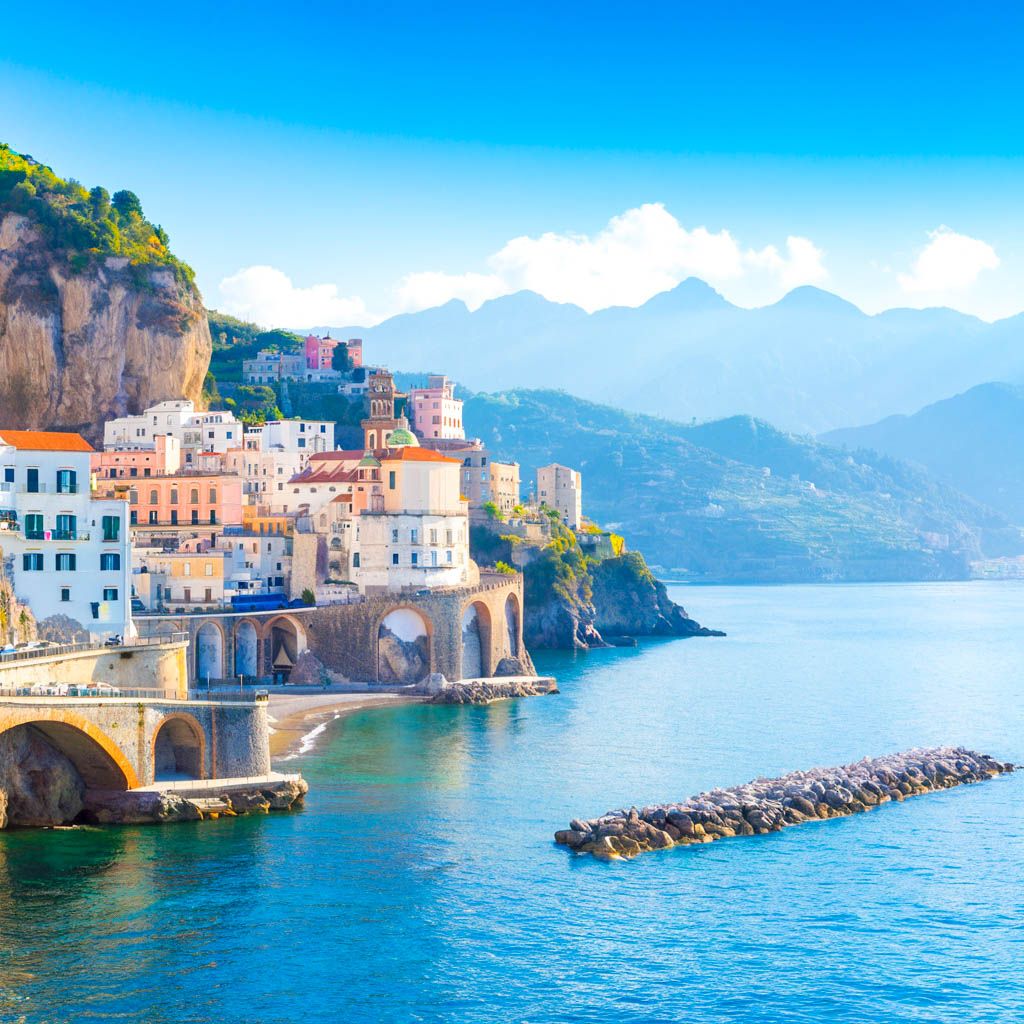} &
\includegraphics[width=\linewidth]{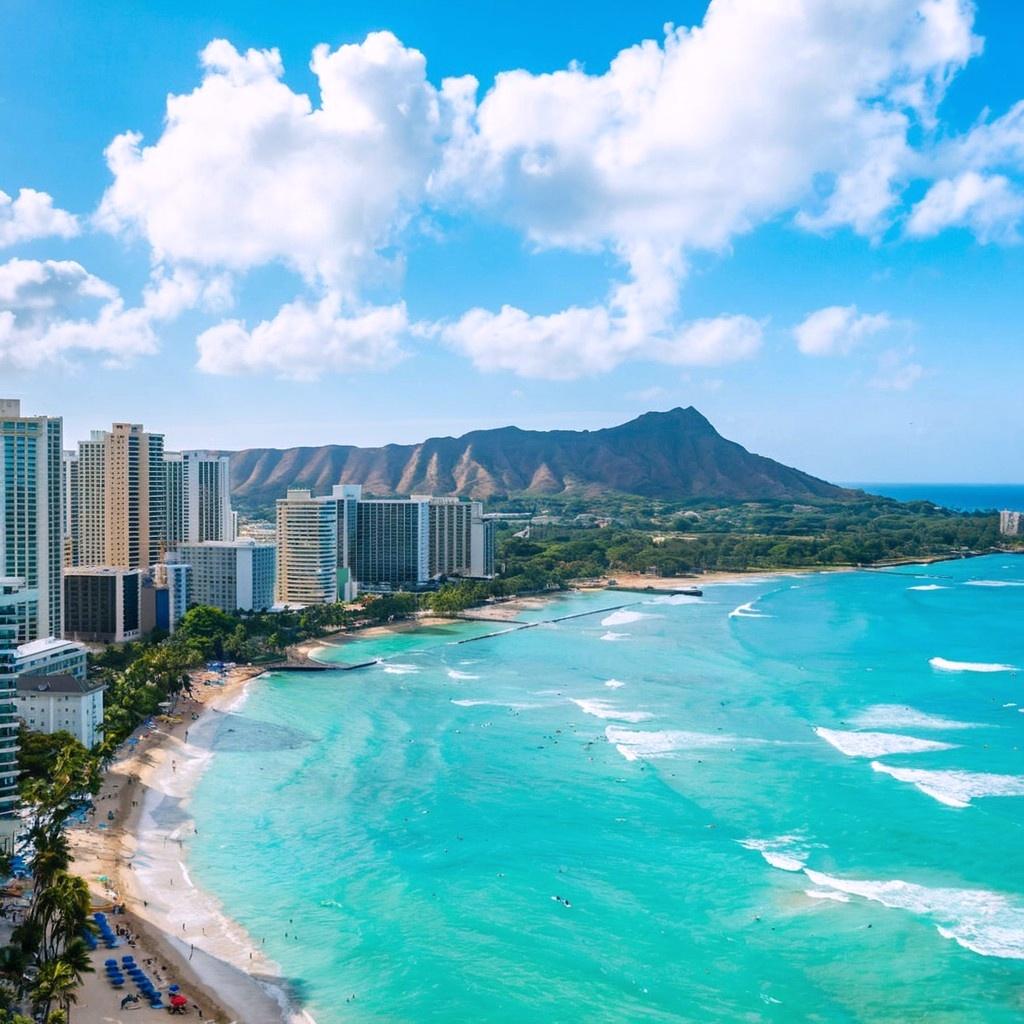} &
\includegraphics[width=\linewidth]{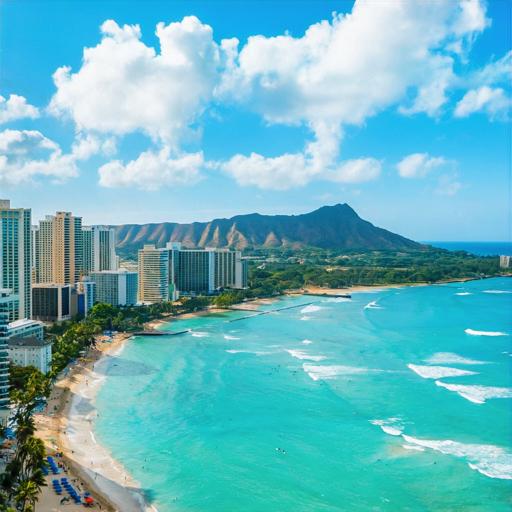} &
\includegraphics[width=\linewidth]{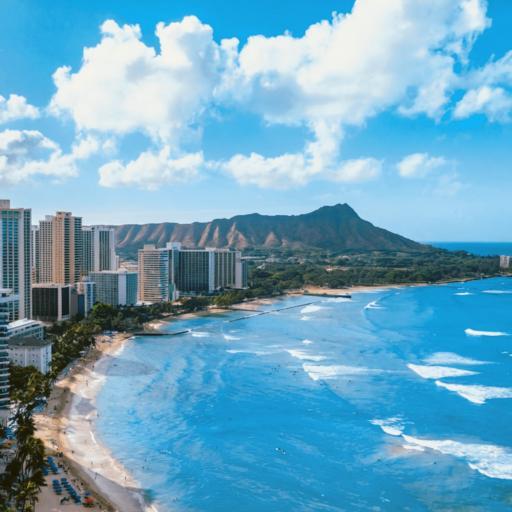}
\\[3pt]

\includegraphics[width=\linewidth]{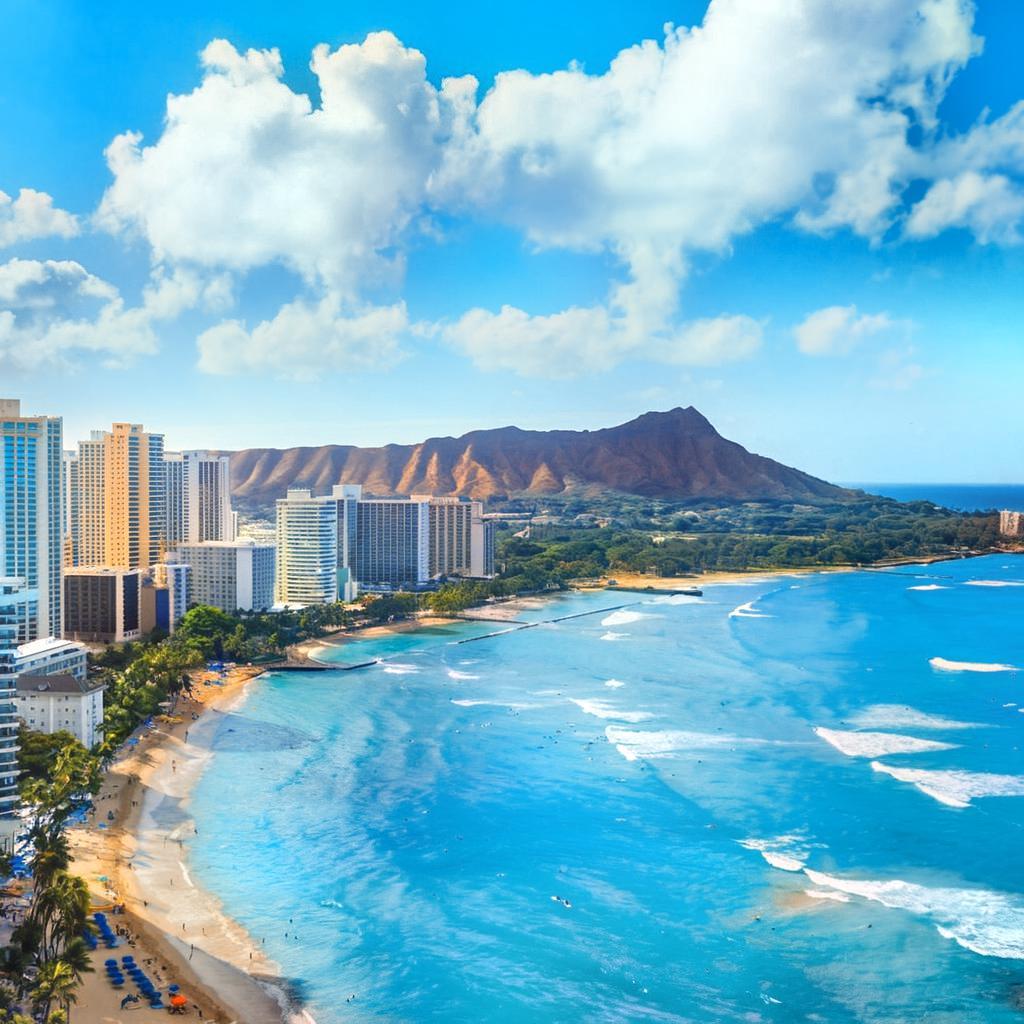} &
\includegraphics[width=\linewidth]{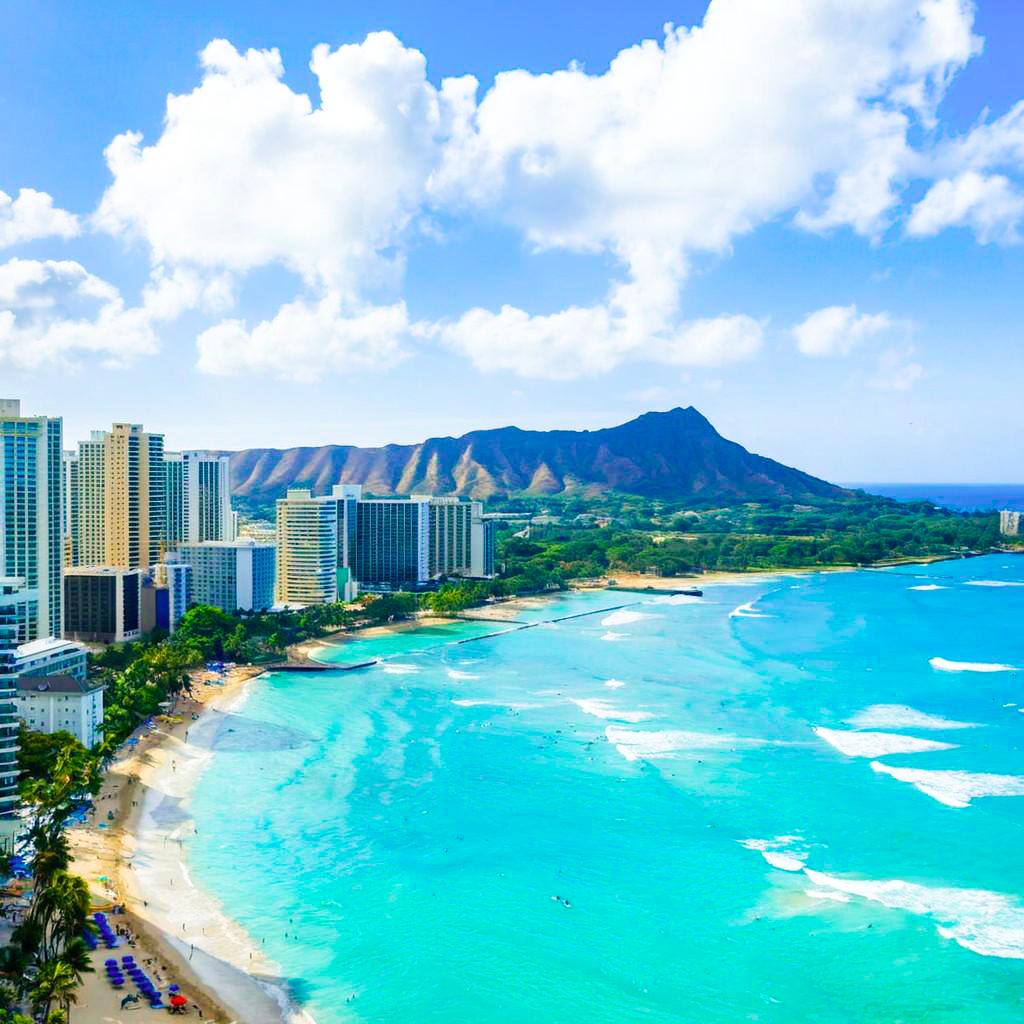} &
\includegraphics[width=\linewidth]{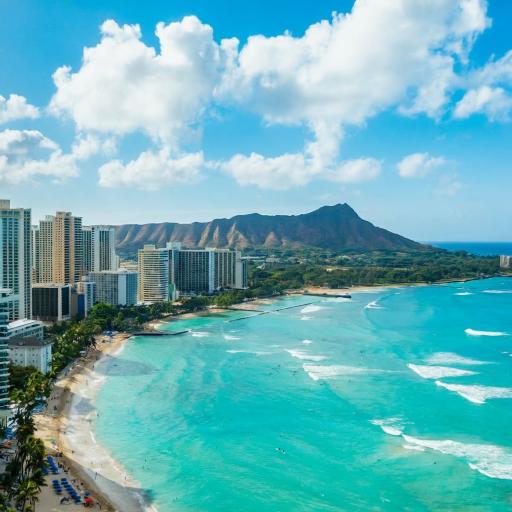} &
\includegraphics[width=\linewidth]{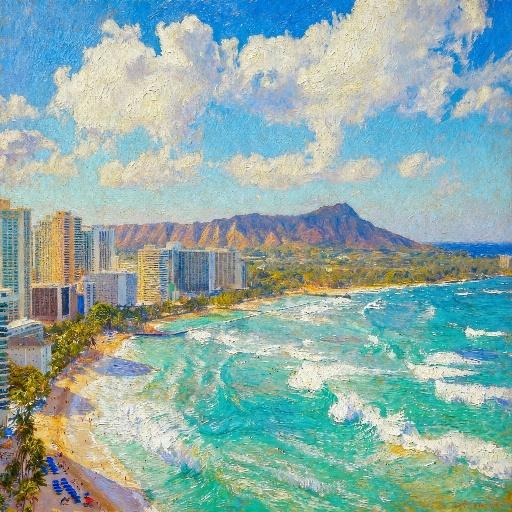} &
\includegraphics[width=\linewidth]{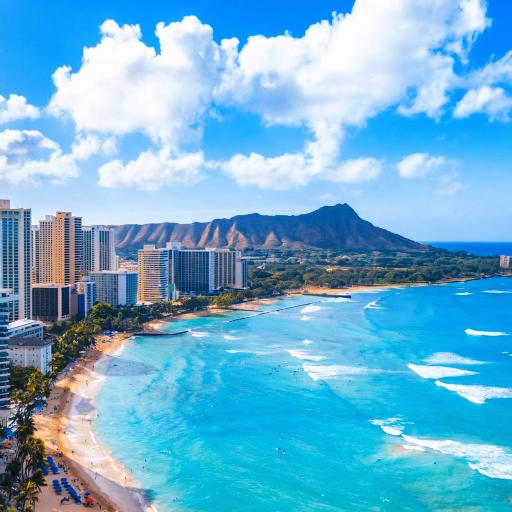}
\\[1pt]

\figlabel{PhotoWCT2~\cite{chiu2022photowct2}} &
\figlabel{PhotoArtAgent~\cite{chen2025photoartagent}} &
\figlabel{NanoBanana~\cite{google2025imagegeneration}} &
\figlabel{Seedream4.0~\cite{seedream2025seedream}} &
\figlabel{Ours}

\end{tabular}}
    \caption{Photorealistic style transfer comparison. Our method can be used for photorealistic style transfer, applying color and tone from a single unpaired reference out of the box.}
    \label{fig:target_only_ref}
\end{figure}

\noindent \textbf{Photorealistic style transfer}\quad Our method naturally supports photorealistic style transfer, in which the color and tonal characteristics of an unpaired style reference are applied to a content image while preserving its structure and semantics~\cite{ho2021deep, ke2023neural, yoo2019photorealistic}. This is achieved by forming a pseudo reference pair, re-using the content image as the input and the style image as the target, and then applying our standard pipeline without any task-specific tuning. The visual results in~\Cref{fig:target_only_ref} and the quantitative results in~\Cref{tab:target_only_scores}, evaluated using metrics from \cite{ke2023neural}, demonstrate that our method achieves superior style transfer performance while faithfully preserving content compared to state-of-the-art Photorealistic style transfer methods \cite{ke2023neural, ho2021deep, chiu2022photowct2, li2025d} as well as generic image editing models \cite{seedream2025seedream, google2025imagegeneration}. This indicates our model's ability to encode retouching style across two semantically different images.

\noindent \textbf{User study} \quad  We further conduct a user study to evaluate both style transfer fidelity and the visual appeal of photorealistic style transfer, criteria that are difficult to fully capture with quantitative metrics due to the lack of ground truth. As shown in~\Cref{tab:user_study}, our method achieves competitive Style Fidelity while outperforming all baselines in Visual Appeal. The study involves 40 participants. For each trial, participants are presented with the reference style image, the content image, and five corresponding stylized results, and are asked to rank the outputs. The evaluated samples are randomly drawn from a pool of 40 examples to increase coverage and reduce sampling bias. Overall, the results indicate that our approach produces perceptually stronger photorealistic stylization, despite not being explicitly trained for the task.


\begin{table*}[htpb]
  \centering
  \caption{Image personalization with few reference samples on the standard MIT-FiveK and PPR10K benchmark datasets. While MSM, StarEnhancer, and our method does not require user-specific training, AdaInt and RSFNet require training on individual user's reference samples.}
  \small
  \begin{tabular}{llccccccccc}
    \toprule
    \multirow{2}{*}{\textbf{Dataset}} & \multirow{2}{*}{\textbf{Method}} & \multicolumn{3}{c}{\textbf{PSNR$\uparrow$}} & \multicolumn{3}{c}{\textbf{SSIM$\uparrow$}} & \multicolumn{3}{c}{\textbf{LPIPS$\downarrow$}} \\
    \cmidrule(lr){3-5} \cmidrule(lr){6-8} \cmidrule(lr){9-11}
    & & \textbf{20} & \textbf{50} & \textbf{100} & \textbf{20} & \textbf{50} & \textbf{100} & \textbf{20} & \textbf{50} & \textbf{100} \\
    \midrule
    \multirow{5}{*}{MIT-FiveK~\cite{fivek}}   
    & MSM \cite{10149499} & 22.13 & 21.79 & 22.05 & 0.807 & 0.806 & 0.807 & 0.171 & 0.174 & 0.171 \\
    & MSM\textsuperscript{\dag} \cite{10149499} & 20.46 & 20.48 & 20.58 & 0.750 & 0.753 & 0.753 & 0.238 & 0.236 & 0.234 \\
    & AdaInt \cite{yang2022adaint} & 20.12 & 20.80 & 22.49 & 0.824 & 0.833 & 0.854 & 0.111 & 0.098 & 0.078  \\ 
    & RSFNet \cite{ouyang2023rsfnet} & 22.52 & 22.87 & 22.66 & 0.862 & 0.866 & 0.86 & 0.072 & 0.069 & 0.072\\ 
    & Ours & \textbf{22.95} & \textbf{23.19} & \textbf{23.34} & \textbf{0.913} & \textbf{0.917} & \textbf{0.920} & \textbf{0.070} & \textbf{0.067} & \textbf{0.067} \\
    \midrule
    \multirow{6}{*}{PPR10K~\cite{liang2021ppr10k}} & StarEnhancer \cite{song2021starenhancer} & 20.54 & 20.57 & 20.64 & 0.880 & 0.878 & 0.882 & 0.085 & 0.086 & 0.085 \\
    & MSM \cite{10149499} & 19.48 & 19.98 & 20.06 & 0.866 & 0.873 & 0.873 & 0.146 & 0.138 & 0.134 \\
    & MSM\textsuperscript{\dag} \cite{10149499} & 20.94 & 21.21 & 21.38 & 0.871 & 0.875 & 0.878 & 0.117 & 0.117 & 0.116 \\
    & AdaInt \cite{yang2022adaint} & 21.28 & 22.08 & \textbf{22.82} & 0.903 & 0.921 & 0.932 & 0.077 & \textbf{0.065} & \textbf{0.058} \\ 
    & RSFNet \cite{ouyang2023rsfnet} & \textbf{22.01} & 22.08 & 22.22 & 0.923 & 0.924 & 0.923 & \textbf{0.069} & 0.067 & 0.065\\ 
    & Ours & 21.75 & \textbf{22.30} & 22.47 & \textbf{0.935} & \textbf{0.941} & \textbf{0.942} & 0.070 & \textbf{0.065} & 0.063 \\
    \bottomrule
  \end{tabular}
  \label{tab:standard_bench_results}
\end{table*}

\noindent \textbf{Retouching-aware feature adaptation}\label{sec:photometric_encoder} \quad We analyze the feature embeddings of the LoRA-adapted SigLIPv2 backbone after training with the asymmetric framework and observe a shift from general semantic representations toward retouching-relevant color and tone representations. In particular, as shown in~\Cref{fig:retrieval}, top-$4$ retrieval using the adapted SigLIPv2 features returns images with similar color and tone characteristics, whereas retrieval using the frozen pretrained SigLIPv2 features is primarily driven by semantic similarity (details of the retrieval implementation are provided in the supplementary material). This observation motivates our use of the adapted SigLIPv2 backbone as the content encoder for multi-reference personalized retouching, since experts typically apply similar retouching styles to images with similar color and tone distributions rather than to images that are merely semantically similar.

\begin{table*}[htpb]
    \centering
    \caption{Ablation study for the \textbf{asymmetric auto-encoder} in RefRetouch. We compare three encoder backbones: \textit{RN-50} (ResNet‑50), \textit{SL-F} (frozen SigLIPv2), and \textit{SL-LoRA} (SigLIPv2 with LoRA adapters). 
    Decoder variants include our proposed conditional \textit{MLP} and the \textit{UNet}-based decoder from~\cite{10149499}. 
    We also evaluate three conditioning strategies: additive (\textit{Add}), adaptive layer normalization (\textit{AdaLN}), and cross-attention (\textit{Cross-attn}). }
    \small
    \begin{tabular}{lcccccccccc}
    \toprule
        \multirow{2}{*}{\textbf{Exp \textnumero}} & \multicolumn{3}{c}{\textbf{Encoder}} & \multicolumn{2}{c}{\textbf{Decoder}} & \multicolumn{3}{c}{\textbf{Cond-Method}} & \multicolumn{2}{c}{\textbf{Metrics}} \\ 
         \cmidrule(lr){2-4} \cmidrule(lr){5-6} \cmidrule(lr){7-9} \cmidrule(lr){10-11}
        & \textit{RN-50} & \textit{SL-F} & \textit{SL-LoRA} 
        & \textit{MLP} & \textit{UNet} 
        & \textit{Add} & \textit{AdaLN} & \textit{Cross-attn} 
        & \textbf{PSNR$\uparrow$} & \textbf{SSIM$\uparrow$} \\
        \midrule
         1 & $\times$ & $\times$ & \checkmark & \checkmark & $\times$ & \checkmark & $\times$ & $\times$ & \textbf{32.40} & \textbf{0.977} \\ 
         2 & \checkmark & $\times$ & $\times$ & \checkmark & $\times$ & \checkmark & $\times$ & $\times$ & 30.00 & 0.971 \\ 
         3 & $\times$ & \checkmark & $\times$ & \checkmark & $\times$ & \checkmark & $\times$ & $\times$ & 23.41 & 0.931 \\
         4 & $\times$ & $\times$ & \checkmark & $\times$ & \checkmark & $\times$ & $\times$ & $\times$ & 29.58 & 0.944 \\ 
         5 & $\times$ & $\times$ & \checkmark & \checkmark & $\times$ & $\times$ & \checkmark & $\times$ & 31.93 & 0.977 \\ 
         6 & $\times$ & $\times$ & \checkmark & \checkmark & $\times$ & $\times$ & $\times$ & \checkmark & 22.66 & 0.880 \\
    \bottomrule
    \end{tabular}
    \label{tab:ablation_study}
\end{table*}

\begin{table}[htpb]
    \centering
    \caption{Image personalization with varying top-$K$, number of relevant references.}
    \setlength{\tabcolsep}{1mm}
    \small
    \begin{tabular}{c|cccc}
    \toprule
    \textbf{Ref} & \textbf{k=1} & \textbf{k=3} & \textbf{k=5} & All \\
    \midrule
    20  & 21.44/0.889 & 22.95/0.913 & 23.37/0.918 & 21.99/0.897 \\
    50  & 21.47/0.899 & 23.19/0.917 & 23.56/0.920 & 21.97/0.898 \\
    100 & 21.94/0.900 & 23.34/0.920 & 23.82/0.923 & 21.94/0.897 \\
    \bottomrule
    \end{tabular}
    \label{tab:top_k_comparison}
\end{table}

\begin{table*}[t]
\centering

\begin{minipage}[t]{0.57\textwidth}
\vspace{0pt}
\centering
\caption{Comparison of content embedding models used for reference
retrieval in RAR. Results are reported in PSNR (dB).}
\label{tab:content_encoder_comparison}

\scriptsize
\setlength{\tabcolsep}{6pt}
\renewcommand{\arraystretch}{1.05}
\begin{tabular}{lcccccc}
    \toprule
    \multirow{2}{*}{\textbf{Encoder}}
    & \multicolumn{3}{c}{\textbf{PPR10K-Groups}}
    & \multicolumn{3}{c}{\textbf{MIT-FiveK}} \\
    \cmidrule(lr){2-4}
    \cmidrule(lr){5-7}
    & \textbf{1} & \textbf{3} & \textbf{6}
    & \textbf{20} & \textbf{50} & \textbf{100} \\
    \midrule
    SigLIPv2 Adapted & 22.51 & 23.92 & 25.27
                     & 22.95 & 23.19 & 23.33 \\
    SigLIPv2 Frozen  & 22.51 & 23.49 & 23.75
                     & 21.02 & 21.38 & 20.99 \\
    Lab Histogram    & 22.51 & 23.94 & 24.66
                     & 21.94 & 21.92 & 21.86 \\
    DINOv2           & 22.51 & 23.53 & 24.49
                     & 21.11 & 20.63 & 20.71 \\
    LUT Decoder      & 22.51 & 23.92 & 25.28
                     & 22.97 & 23.21 & 23.34 \\
    \bottomrule
\end{tabular}
\end{minipage}
\hfill
\begin{minipage}[t]{0.40\textwidth}
\vspace{0pt}
\centering
\caption{User study on photorealistic style transfer. T1 and T2
denote Top-1 and Top-2 percentages.}
\label{tab:user_study}

\scriptsize
\setlength{\tabcolsep}{3pt}
\renewcommand{\arraystretch}{1.05}
\begin{tabular}{lcccc}
    \toprule
    \textbf{Method}
    & \multicolumn{2}{c}{\textbf{Style Fid.}}
    & \multicolumn{2}{c}{\textbf{Vis. Appeal}} \\
    \cmidrule(lr){2-3}
    \cmidrule(lr){4-5}
    & T1 (\%) & T2 (\%) & T1 (\%) & T2 (\%) \\
    \midrule
    D-LUT      & 9.8  & 23.5 & 13.7 & 49.0 \\
    DeepPreset & 15.7 & 29.4 & 37.3 & 66.7 \\
    NanoBanana & 7.8  & 17.6 & 5.9  & 29.4 \\
    PhotoWCT2  & 41.2 & 66.7 & 15.7 & 35.3 \\
    Ours       & 39.2 & 80.4 & 41.2 & 56.9 \\
    \bottomrule
\end{tabular}
\end{minipage}

\end{table*}

\subsection{Ablation Study}\label{sec:ablations}

To validate the architectural design and evaluate the contribution of individual components in the asymmetric auto-encoder of \textit{RefRetouch}, we conduct ablation studies varying the encoder backbone, decoder architecture, and conditioning method, as summarized in~\Cref{tab:ablation_study}. All models are trained for 10,000 iterations on a subset of the training data with a batch size of 8. Reconstruction fidelity is evaluated on a held-out validation set using PSNR and SSIM metrics.

\noindent \textbf{Encoder backbone} \quad In experiments 1–3, we observe that using a SigLIPv2 encoder with trainable LoRA adapters significantly outperforms both the ResNet-50 and frozen SigLIPv2 variants. This highlights that strong pretrained features alone are insufficient; adaptation to the retouching task is essential for effective retouching style extraction.

\noindent \textbf{MLP decoder} \quad Comparing Experiments 1 and 4, we find that the proposed lightweight color-space MLP decoder achieves better reconstruction performance than the UNet-based decoder used in \cite{10149499}, highlighting the effectiveness of our asymmetric design. This result supports our hypothesis that a simple per-pixel decoder forces the encoder to capture content-disentangled, retouching-relevant features, enabling faithful reconstruction of the enhanced output from the input. 

\noindent \textbf{Conditioning method}\quad Experiments 1, 5, and 6 further demonstrate that simple additive conditioning injection outperforms adaptive layer norm and cross-attention alternatives. For conditioning the UNet decoder with the retouching style latent, we adopt the approach in \cite{10149499}.

\noindent \textbf{Retrieval-augmented retouching}\quad \Cref{tab:top_k_comparison} shows the image personalization performance on MIT-FiveK, varying the number of relevant references used to compute the retouching style, as in~\Cref{eq:rar}. Performance improves with more relevant references, demonstrating the model's ability to utilize multiple inputs. However, using all the references results in poor performance, highlighting the importance of selecting relevant samples and emphasizing the effectiveness of our retrieval-augmented retouching approach.

\noindent \textbf{Content embedding model}\quad
We compare different content embedding models for representing the query and reference images in the RAR module. The results in~\Cref{tab:content_encoder_comparison} provide the retouching performance on the multi-reference PPR10K-Groups and MIT-FiveK benchmarks. Overall, the LoRA-adapted SigLIPv2 backbone consistently achieves the best performance, outperforming generic semantic encoders such as frozen SigLIPv2 and DINOv2, while also surpassing the hand-crafted Lab histogram descriptor. These results suggest that the asymmetric auto-encoding framework adapts the representation toward retouching-relevant photometric attributes, enabling more effective retrieval of reference images with similar color and tone characteristics.

\noindent\textbf{Limitations.} Our method currently performs only global retouching, applying a uniform transformation across the entire image. As a result, it cannot selectively adjust specific regions or handle spatially varying edits. In future work, we aim to extend the framework to support region-aware, locally adaptive retouching by incorporating spatial decomposition or region-level conditioning following the approaches in \cite{zeng2023region, lin2025jarvisart}.

\section{Conclusion}

We proposed a generic personalized image retouching framework that adapts to individual user’s image retouching style from only a handful of reference pairs without test-time fine-tuning. Experiments across three benchmarks of increasing complexity show that our proposed method consistently outperforms existing generic image retouching personalization methods. In addition, the same pipeline can be used out of the box for photorealistic style transfer, highlighting its versatility and generalization across tasks.

{
    \small
    \bibliographystyle{ieeenat_fullname}
    \bibliography{main}

@String(ECCV= {Eur. Conf. Comput. Vis.})

@String(TOG= {ACM Trans. Graph.})

@String(AAAI = {AAAI})

@String(ECCV  = {ECCV})

@String(TOG   = {ACM TOG})

@article{zeng2020learning,
  title={Learning image-adaptive 3d lookup tables for high performance photo enhancement in real-time},
  author={Zeng, Hui and Cai, Jianrui and Li, Lida and Cao, Zisheng and Zhang, Lei},
  journal={IEEE Transactions on Pattern Analysis and Machine Intelligence},
  volume={44},
  number={4},
  pages={2058--2073},
  year={2020},
  publisher={IEEE}
}

@inproceedings{yang2022adaint,
  title={Adaint: Learning adaptive intervals for 3d lookup tables on real-time image enhancement},
  author={Yang, Canqian and Jin, Meiguang and Jia, Xu and Xu, Yi and Chen, Ying},
  booktitle={Proceedings of the IEEE/CVF Conference on Computer Vision and Pattern Recognition},
  pages={17522--17531},
  year={2022}
}

@inproceedings{ouyang2023rsfnet,
  title={Rsfnet: A white-box image retouching approach using region-specific color filters},
  author={Ouyang, Wenqi and Dong, Yi and Kang, Xiaoyang and Ren, Peiran and Xu, Xin and Xie, Xuansong},
  booktitle={Proceedings of the IEEE/CVF International Conference on Computer Vision},
  pages={12160--12169},
  year={2023}
}

@article{gharbi2017deep,
  title={Deep bilateral learning for real-time image enhancement},
  author={Gharbi, Micha{\"e}l and Chen, Jiawen and Barron, Jonathan T and Hasinoff, Samuel W and Durand, Fr{\'e}do},
  journal={ACM Transactions on Graphics (TOG)},
  volume={36},
  number={4},
  pages={1--12},
  year={2017},
  publisher={ACM New York, NY, USA}
}

@inproceedings{conde2024nilut,
  title={Nilut: Conditional neural implicit 3d lookup tables for image enhancement},
  author={Conde, Marcos V and Vazquez-Corral, Javier and Brown, Michael S and Timofte, Radu},
  booktitle={Proceedings of the AAAI Conference on Artificial Intelligence},
  year={2024}
}

@inproceedings{he2020conditional,
  title={Conditional sequential modulation for efficient global image retouching},
  author={He, Jingwen and Liu, Yihao and Qiao, Yu and Dong, Chao},
  booktitle={Computer Vision--ECCV 2020: 16th European Conference, Glasgow, UK, August 23--28, 2020, Proceedings, Part XIII 16},
  pages={679--695},
  year={2020},
  organization={Springer}
}

@inproceedings{moran2020deeplpf,
  title={Deeplpf: Deep local parametric filters for image enhancement},
  author={Moran, Sean and Marza, Pierre and McDonagh, Steven and Parisot, Sarah and Slabaugh, Gregory},
  booktitle={Proceedings of the IEEE/CVF conference on computer vision and pattern recognition},
  pages={12826--12835},
  year={2020}
}

@inproceedings{wang2019underexposed,
  title={Underexposed photo enhancement using deep illumination estimation},
  author={Wang, Ruixing and Zhang, Qing and Fu, Chi-Wing and Shen, Xiaoyong and Zheng, Wei-Shi and Jia, Jiaya},
  booktitle={Proceedings of the IEEE/CVF conference on computer vision and pattern recognition},
  pages={6849--6857},
  year={2019}
}

@inproceedings{chen2018deep,
  title={Deep photo enhancer: Unpaired learning for image enhancement from photographs with gans},
  author={Chen, Yu-Sheng and Wang, Yu-Ching and Kao, Man-Hsin and Chuang, Yung-Yu},
  booktitle={Proceedings of the IEEE conference on computer vision and pattern recognition},
  pages={6306--6314},
  year={2018}
}

@inproceedings{isola2017image,
  title={Image-to-image translation with conditional adversarial networks},
  author={Isola, Phillip and Zhu, Jun-Yan and Zhou, Tinghui and Efros, Alexei A},
  booktitle={Proceedings of the IEEE conference on computer vision and pattern recognition},
  pages={1125--1134},
  year={2017}
}

@inproceedings{wang2022neural,
  title={Neural color operators for sequential image retouching},
  author={Wang, Yili and Li, Xin and Xu, Kun and He, Dongliang and Zhang, Qi and Li, Fu and Ding, Errui},
  booktitle={European Conference on Computer Vision},
  pages={38--55},
  year={2022},
  organization={Springer}
}

@article{hu2018exposure,
  title={Exposure: A white-box photo post-processing framework},
  author={Hu, Yuanming and He, Hao and Xu, Chenxi and Wang, Baoyuan and Lin, Stephen},
  journal={ACM Transactions on Graphics (TOG)},
  volume={37},
  number={2},
  pages={1--17},
  year={2018},
  publisher={ACM New York, NY, USA}
}

@inproceedings{shi2021learning,
  title={Learning by planning: Language-guided global image editing},
  author={Shi, Jing and Xu, Ning and Xu, Yihang and Bui, Trung and Dernoncourt, Franck and Xu, Chenliang},
  booktitle={Proceedings of the IEEE/CVF Conference on Computer Vision and Pattern Recognition},
  pages={13590--13599},
  year={2021}
}

@inproceedings{kim2020pienet,
  title={PieNet: Personalized image enhancement network},
  author={Kim, Han-Ul and Koh, Young Jun and Kim, Chang-Su},
  booktitle={Computer Vision--ECCV 2020: 16th European Conference, Glasgow, UK, August 23--28, 2020, Proceedings, Part XXX 16},
  pages={374--390},
  year={2020},
  organization={Springer}
}

@inproceedings{song2021starenhancer,
  title={Starenhancer: Learning real-time and style-aware image enhancement},
  author={Song, Yuda and Qian, Hui and Du, Xin},
  booktitle={Proceedings of the IEEE/CVF International Conference on Computer Vision},
  pages={4126--4135},
  year={2021}
}

@ARTICLE{10149499,
  author={Kosugi, Satoshi and Yamasaki, Toshihiko},
  journal={IEEE Transactions on Circuits and Systems for Video Technology}, 
  title={Personalized Image Enhancement Featuring Masked Style Modeling}, 
  year={2024},
  volume={34},
  number={1},
  pages={140-152},
  doi={10.1109/TCSVT.2023.3285765}
}

@article{wang2023learning,
  title={Learning diverse tone styles for image retouching},
  author={Wang, Haolin and Zhang, Jiawei and Liu, Ming and Wu, Xiaohe and Zuo, Wangmeng},
  journal={IEEE Transactions on Image Processing},
  volume={33},
  pages={310--321},
  year={2023},
  publisher={IEEE}
}

@article{kim2025oneta,
  title={Oneta: Multi-Style Image Enhancement Using Eigentransformation Functions},
  author={Kim, Jiwon and Hwang, Soohyun and Kim, Dong-O and Han, Changsu and Park, Min Kyu and Kim, Chang-Su},
  journal={arXiv preprint arXiv:2506.23547},
  year={2025}
}

@inproceedings{duan2025diffretouch,
  title={DiffRetouch: Using Diffusion to Retouch on the Shoulder of Experts},
  author={Duan, Zheng-Peng and Zhang, Jiawei and Lin, Zheng and Jin, Xin and Wang, XunDong and Zou, Dongqing and Guo, Chun-Le and Li, Chongyi},
  booktitle={Proceedings of the AAAI Conference on Artificial Intelligence},
  year={2025}
}

@inproceedings{ke2023neural,
  title={Neural preset for color style transfer},
  author={Ke, Zhanghan and Liu, Yuhao and Zhu, Lei and Zhao, Nanxuan and Lau, Rynson WH},
  booktitle={Proceedings of the IEEE/CVF conference on computer vision and pattern recognition},
  pages={14173--14182},
  year={2023}
}

@inproceedings{ho2021deep,
  title={Deep preset: Blending and retouching photos with color style transfer},
  author={Ho, Man M and Zhou, Jinjia},
  booktitle={Proceedings of the IEEE/CVF Winter Conference on Applications of Computer Vision},
  pages={2113--2121},
  year={2021}
}

@inproceedings{yoo2019photorealistic,
  title={Photorealistic style transfer via wavelet transforms},
  author={Yoo, Jaejun and Uh, Youngjung and Chun, Sanghyuk and Kang, Byeongkyu and Ha, Jung-Woo},
  booktitle={Proceedings of the IEEE/CVF international conference on computer vision},
  pages={9036--9045},
  year={2019}
}

@inproceedings{li2018closed,
  title={A closed-form solution to photorealistic image stylization},
  author={Li, Yijun and Liu, Ming-Yu and Li, Xueting and Yang, Ming-Hsuan and Kautz, Jan},
  booktitle={Proceedings of the European conference on computer vision (ECCV)},
  pages={453--468},
  year={2018}
}

@inproceedings{chiu2022photowct2,
  title={Photowct2: Compact autoencoder for photorealistic style transfer resulting from blockwise training and skip connections of high-frequency residuals},
  author={Chiu, Tai-Yin and Gurari, Danna},
  booktitle={Proceedings of the IEEE/CVF winter conference on applications of computer vision},
  pages={2868--2877},
  year={2022}
}

@inproceedings{li2025d,
  title={D-LUT: Photorealistic Style Transfer via Diffusion Process},
  author={Li, Mujing and Wang, Guanjie and Zhang, Xingguang and Liao, Qifeng and Xiao, Chenxi},
  booktitle={2025 IEEE/CVF Winter Conference on Applications of Computer Vision (WACV)},
  pages={9206--9214},
  year={2025},
  organization={IEEE}
}

@article{elezabi2024inretouch,
  title={INRetouch: Context Aware Implicit Neural Representation for Photography Retouching},
  author={Elezabi, Omar and Conde, Marcos V and Wu, Zongwei and Timofte, Radu},
  journal={arXiv preprint arXiv:2412.03848},
  year={2024}
}

@article{wu2024goal,
  title={Goal Conditioned Reinforcement Learning for Photo Finishing Tuning},
  author={Wu, Jiarui and Wang, Yujin and Li, Lingen and Zhang, Fan and Xue, Tianfan},
  journal={Advances in Neural Information Processing Systems},
  volume={37},
  pages={46294--46318},
  year={2024}
}

@article{tseng2022neural,
  title={Neural Photo-Finishing.},
  author={Tseng, Ethan and Zhang, Yuxuan and Jebe, Lars and Zhang, Xuaner and Xia, Zhihao and Fan, Yifei and Heide, Felix and Chen, Jiawen},
  journal={ACM Trans. Graph.},
  volume={41},
  number={6},
  pages={238--1},
  year={2022}
}

@article{li2025visualcloze,
  title={VisualCloze: A Universal Image Generation Framework via Visual In-Context Learning},
  author={Li, Zhong-Yu and Du, Ruoyi and Yan, Juncheng and Zhuo, Le and Li, Zhen and Gao, Peng and Ma, Zhanyu and Cheng, Ming-Ming},
  journal={arXiv preprint arXiv:2504.07960},
  year={2025}
}

@article{chen2025edit,
  title={Edit Transfer: Learning Image Editing via Vision In-Context Relations},
  author={Chen, Lan and Mao, Qi and Gu, Yuchao and Shou, Mike Zheng},
  journal={arXiv preprint arXiv:2503.13327},
  year={2025}
}

@article{bar2022visual,
  title={Visual prompting via image inpainting},
  author={Bar, Amir and Gandelsman, Yossi and Darrell, Trevor and Globerson, Amir and Efros, Alexei},
  journal={Advances in Neural Information Processing Systems},
  volume={35},
  pages={25005--25017},
  year={2022}
}

@inproceedings{wang2023images,
  title={Images speak in images: A generalist painter for in-context visual learning},
  author={Wang, Xinlong and Wang, Wen and Cao, Yue and Shen, Chunhua and Huang, Tiejun},
  booktitle={Proceedings of the IEEE/CVF Conference on Computer Vision and Pattern Recognition},
  pages={6830--6839},
  year={2023}
}

@inproceedings{fivek,
	author = "Vladimir Bychkovsky and Sylvain Paris and Eric Chan and Fr{\'e}do Durand",
	title = "Learning Photographic Global Tonal Adjustment with a Database of Input / Output Image Pairs",
	booktitle = "The Twenty-Fourth IEEE Conference on Computer Vision and Pattern Recognition",
	year = "2011"
}

@inproceedings{liang2021ppr10k,
  title={Ppr10k: A large-scale portrait photo retouching dataset with human-region mask and group-level consistency},
  author={Liang, Jie and Zeng, Hui and Cui, Miaomiao and Xie, Xuansong and Zhang, Lei},
  booktitle={Proceedings of the IEEE/CVF Conference on Computer Vision and Pattern Recognition},
  pages={653--661},
  year={2021}
}

@article{schuhmann2022laion,
  title={Laion-5b: An open large-scale dataset for training next generation image-text models},
  author={Schuhmann, Christoph and Beaumont, Romain and Vencu, Richard and Gordon, Cade and Wightman, Ross and Cherti, Mehdi and Coombes, Theo and Katta, Aarush and Mullis, Clayton and Wortsman, Mitchell and others},
  journal={Advances in neural information processing systems},
  volume={35},
  pages={25278--25294},
  year={2022}
}

@article{tschannen2025siglip,
  title={Siglip 2: Multilingual vision-language encoders with improved semantic understanding, localization, and dense features},
  author={Tschannen, Michael and Gritsenko, Alexey and Wang, Xiao and Naeem, Muhammad Ferjad and Alabdulmohsin, Ibrahim and Parthasarathy, Nikhil and Evans, Talfan and Beyer, Lucas and Xia, Ye and Mustafa, Basil and others},
  journal={arXiv preprint arXiv:2502.14786},
  year={2025}
}

@misc{adobe_lightroom,
  author       = {{Adobe Inc.}},
  title        = {Adobe lightroom community presets},
  url          = {https://lightroom.adobe.com/learn/discover},
  year         = {2025},
  organization = {Adobe Inc.},
  note         = {Accessed 2025‑05‑11}
}

@inproceedings{hoffer2015deep,
  title={Deep metric learning using triplet network},
  author={Hoffer, Elad and Ailon, Nir},
  booktitle={International workshop on similarity-based pattern recognition},
  pages={84--92},
  year={2015},
  organization={Springer}
}

@inproceedings{peebles2023scalable,
  title={Scalable diffusion models with transformers},
  author={Peebles, William and Xie, Saining},
  booktitle={Proceedings of the IEEE/CVF international conference on computer vision},
  pages={4195--4205},
  year={2023}
}

@article{chen2025photoartagent,
  title={PhotoArtAgent: Intelligent Photo Retouching with Language Model-Based Artist Agents},
  author={Chen, Haoyu and Tao, Keda and Wang, Yizao and Wang, Xinlei and Zhu, Lei and Gu, Jinjin},
  journal={arXiv preprint arXiv:2505.23130},
  year={2025}
}

@article{dutt2025monetgpt,
  title={MonetGPT: Solving Puzzles Enhances MLLMs' Image Retouching Skills},
  author={Dutt, Niladri Shekhar and Ceylan, Duygu and Mitra, Niloy J},
  journal={arXiv preprint arXiv:2505.06176},
  year={2025}
}

@article{lin2025jarvisart,
  title={JarvisArt: Liberating Human Artistic Creativity via an Intelligent Photo Retouching Agent},
  author={Lin, Yunlong and Lin, Zixu and Lin, Kunjie and Bai, Jinbin and Pan, Panwang and Li, Chenxin and Chen, Haoyu and Wang, Zhongdao and Ding, Xinghao and Li, Wenbo and others},
  journal={arXiv preprint arXiv:2506.17612},
  year={2025}
}

@article{seedream2025seedream,
  title={Seedream 4.0: Toward next-generation multimodal image generation},
  author={Seedream, Team and Chen, Yunpeng and Gao, Yu and Gong, Lixue and Guo, Meng and Guo, Qiushan and Guo, Zhiyao and Hou, Xiaoxia and Huang, Weilin and Huang, Yixuan and others},
  journal={arXiv preprint arXiv:2509.20427},
  year={2025}
}

@misc{google2025imagegeneration,
  author       = {Google AI},
  title        = {Image generation with Gemini (aka Nano Banana)},
  howpublished = {\url{https://ai.google.dev/gemini-api/docs/image-generation}},
  year         = {2025},
  note         = {Accessed: 2025-11-12}
}

@article{zeng2023region,
  title={Region-aware portrait retouching with sparse interactive guidance},
  author={Zeng, Huimin and Huang, Jie and Li, Jiacheng and Xiong, Zhiwei},
  journal={IEEE Transactions on Multimedia},
  volume={26},
  pages={127--140},
  year={2023},
  publisher={IEEE}
}

@inproceedings{kim2024image,
  title={Image-adaptive 3d lookup tables for real-time image enhancement with bilateral grids},
  author={Kim, Wontae and Cho, Nam Ik},
  booktitle={European Conference on Computer Vision},
  pages={91--108},
  year={2024},
  organization={Springer}
}

@inproceedings{serrano2024namedcurves,
  title={Namedcurves: Learned image enhancement via color naming},
  author={Serrano-Lozano, David and Herranz, Luis and Brown, Michael S and Vazquez-Corral, Javier},
  booktitle={European Conference on Computer Vision},
  pages={92--108},
  year={2024},
  organization={Springer}
}

@inproceedings{kim2025lightweight,
  title={Lightweight and Fast Real-time Image Enhancement via Decomposition of the Spatial-aware Lookup Tables},
  author={Kim, Wontae and Lee, Keuntek and Cho, Nam Ik},
  booktitle={Proceedings of the IEEE/CVF International Conference on Computer Vision},
  pages={11895--11905},
  year={2025}
}
}

\clearpage
\maketitlesupplementary

\renewcommand{\thesection}{\Alph{section}}
\renewcommand{\thefigure}{\Alph{figure}}
\renewcommand{\thetable}{\Alph{table}}

\section{Visually Consistent Image Retouching Dataset}\label{sec:dataset}

Training a generic few-shot image retouching model \cite{10149499, song2021starenhancer} requires a large and diverse set of retouching styles for meta-learning. However, existing datasets such as MIT-FiveK \cite{fivek} and PPR10K \cite{liang2021ppr10k} include images retouched by only five and three experts, respectively. In addition to this limited stylistic diversity, the human edits in these datasets are often inconsistent across images, making them unsuitable for generic retouching tasks. The method in \cite{10149499} attempts to expand stylistic variety by collecting human-retouched images from Flickr, but these samples contain only the retouched targets and lack their corresponding original inputs. To overcome this, the authors train a degradation model on the paired MIT-FiveK dataset to approximate pseudo-inputs by mapping the targets back to their presumed originals. Yet, because MIT-FiveK primarily consists of underexposed images, the resulting pseudo-inputs are still limited in diversity and fail to generalize to broader content.

\begin{figure}[t] 
    \centering
    \includegraphics[width=0.9\linewidth]{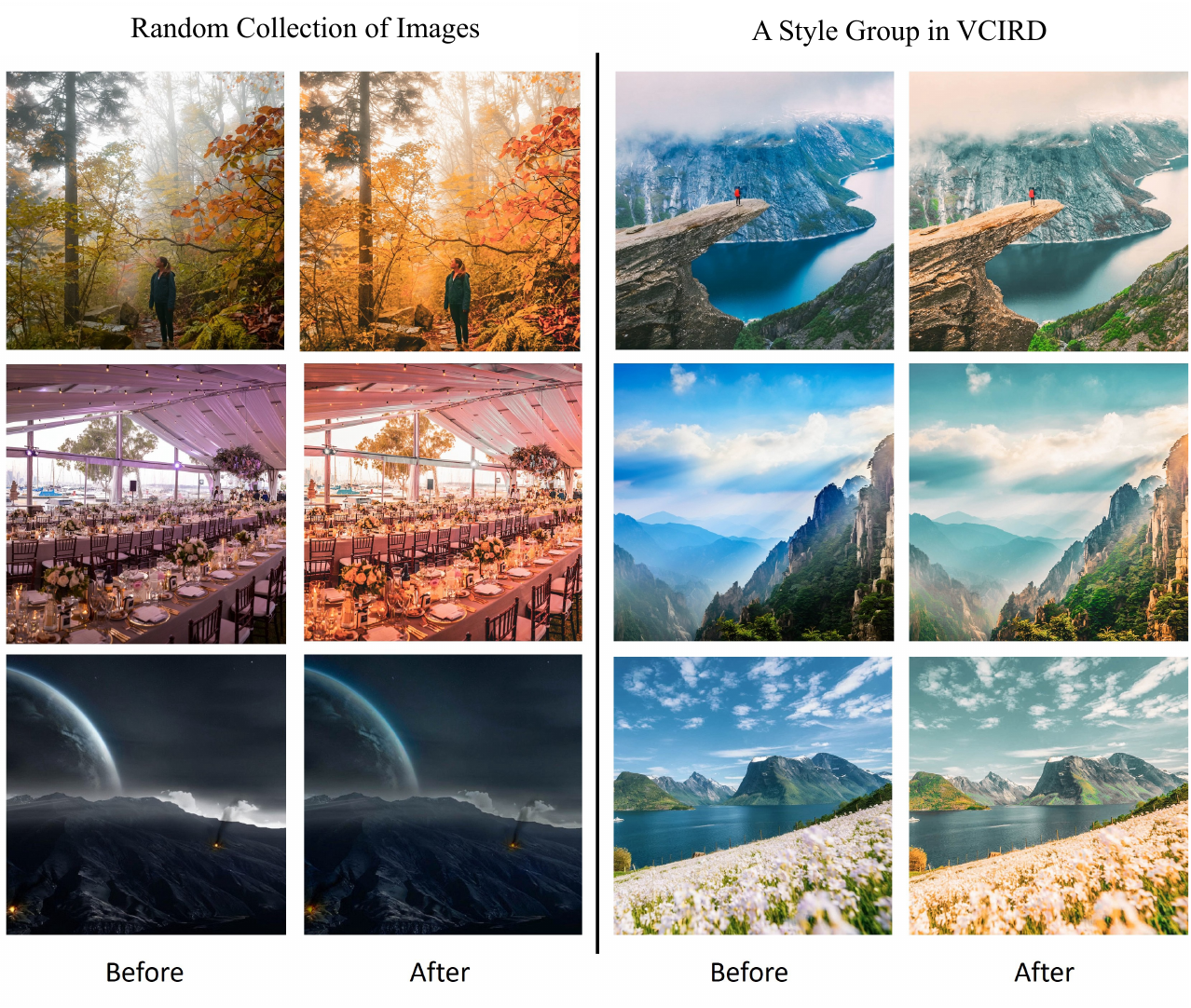}
    \caption{ Effects of applying a single preset to different types of images. The preset produces varying results depending on the input’s context, primarily its semantics and color distribution. Left: Applying the preset to a random set of images leads to inconsistent styles. Right: A style group from our dataset, applying the preset to images with similar context results in perceptually consistent style groups, enabling effective few-shot image retouching. }
    \label{fig:dataset}
\end{figure}

Several works \cite{ke2023neural, ho2021deep, elezabi2024inretouch} construct large-scale retouching datasets by applying Adobe Lightroom presets \cite{adobe_lightroom} to collections of images. While this preset-based strategy increases style diversity, applying presets indiscriminately across random images can lead to visually inconsistent results, since the same preset affects different images differently depending on their context (see~\Cref{fig:dataset}). To address this, we construct a visually consistent retouching dataset by applying carefully curated Lightroom presets to images grouped by coherent semantics and color distributions. To ensure diverse input contexts, we sample high-resolution images from the LAION dataset \cite{schuhmann2022laion}, covering a wide range of colors, tones, and semantic content.

To reduce style variance within each group, we organize the images using a two-stage clustering approach. First, we categorize them into 30 distinct semantic groups using the vision-language embeddings from the pretrained SigLIPv2 model \cite{tschannen2025siglip}. Then, within each semantic category, we apply K-means clustering based on the 2D histogram of the ab channels in the Lab color space, ensuring that images within a sub-cluster share similar color characteristics. We choose the number of clusters to balance granularity, keeping them small enough to maintain color consistency, yet large enough to preserve diversity. This results in a total of \textbf{2,373} distinct groups.

Once images are grouped by both semantic content and color distribution, we apply carefully curated Adobe Lightroom presets to each group. We curate a set of 830 Lightroom presets by selecting sRGB images to avoid clipping artifacts and filtering out presets that involve masks, vignetting, geometric transformations (e.g., cropping), or texture enhancements. These types of edits cannot be easily transferred across different images to build consistent retouching styles. Therefore, we limit our scope to presets that perform only color and tone transformations.

To avoid contextually mismatched edits that may yield degraded results, we pre-filter presets applied to images in a particular group by matching the semantic labels of the group to the images that the presets are originally applied to using CLIP embedding. Through this process, we associate 40 suitable presets with each image group, resulting in a consistent image retouching dataset comprising \textbf{94,400} stylized image groups. As illustrated in~\Cref{fig:dataset}, applying the same preset to a semantically and chromatically coherent group produces perceptually consistent results across all images. This consistency is crucial for meta-learning image retouching, as both reference and query images can be reliably sampled from the same style group. In contrast to datasets where presets are applied to randomly selected images, our approach ensures that reference pairs offer informative and contextually relevant guidance for retouching the query images.

\section{Benchmark Datasets}

We evaluate image retouching personalization across three increasingly challenging settings: (1) single-style retouching, (2) multi-style retouching with consistent edits, and (3) multi-style retouching with inconsistent, implicitly defined edits.

\noindent\textbf{Visually Consistent Image Retouching Benchmark (VCIRB).}
To evaluate generic image retouching methods under a consistent retouching setup where both the reference and test pairs follow the same retouching transformation, we construct the VCIRB benchmark. We hold out 25 groups of images, each edited by a distinct Lightroom preset, from the dataset described in~\Cref{sec:dataset}. Both the images and the corresponding presets are excluded from training to ensure unbiased evaluation of models' generalization to unseen images and styles. For each held-out group, one or more reference pairs are then sampled to be used as exemplars to retouch the remaining images in the group. Then, image retouching model's performance in adapting the retouching style provided by the reference pairs to the testing images is evaluated by comparing the model's output to the ground truth targets. 

\noindent\textbf{PPR10K-Groups.}
The performance of a model under a consistent retouching setup (as in VCIRB) reflects its capacity to understand and replicate the retouching transformation exemplified in the given reference images. However, in practice, users often seek to contextually retouch images by providing a set of diverse reference examples. For example, in event or portrait photography, users may first retouch a few representative images and then use these as references to batch-retouch a large collection of images captured under similar conditions. In these scenarios, images that share similar input contexts, such as scenery, tone, or color attributes, are expected to exhibit similar retouching style adjustments. To evaluate a model’s performance in such context-aware scenarios, we employ the PPR10K dataset \cite{liang2021ppr10k}. This dataset groups portrait images of subjects captured across varying scenes, where professional experts have applied consistent retouching styles within each group. This inherent style consistency guarantees that only a few reference samples within a group are sufficient to convey the user’s retouching preferences, while the variation in scene context (e.g., backgrounds, lighting, tones) naturally tests a model’s ability to adapt those preferences to different input conditions. We select 21 PPR10K groups retouched by Expert A, each containing more than 12 images.

\noindent\textbf{MIT-FiveK and PPR10K.}
We also evaluated image retouching adaptation using the standard MIT-FiveK and PPR10K datasets, which focus on retouching personalization based on a user’s historical edits and stylistic preferences. These datasets contain images retouched by multiple experts, each forming a distinct user style, and are divided into predefined training and testing splits. The training split is used to learn or adapt to a particular user’s retouching style, while the corresponding test split evaluates the model’s personalization performance on unseen images. Unlike the first two setups, the retouching styles in this configuration are often more complex and subtle, as the datasets include a diverse range of image contents and exhibit notable retouching inconsistencies. The same user’s stylistic adjustments may vary significantly across images, even when they share similar visual attributes. Consequently, personalization in this setting requires a larger number of reference samples to effectively capture the user’s underlying retouching patterns. In our experiments, we sampled a few representative reference images from the training split (see~\Cref{sec:ref_sample} for details), while using the same test split for evaluation as defined in the original benchmark \cite{yang2022adaint}.


\begin{algorithm}[t]
\caption{Diverse and Representative Reference Set Sampling}
\label{alg:diverse_ref_sampling}
\begin{algorithmic}[1]
\Require Feature set $\mathcal{H} = \{\mathbf{h}_1, \mathbf{h}_2, \ldots, \mathbf{h}_n\}$, sample size $k$, distance function $d(\cdot, \cdot)$
\Ensure Selected indices $\mathcal{S} \subset \{1, 2, \ldots, n\}$ with $|\mathcal{S}| = k$

\State $\mathcal{S}_0 \gets \emptyset$, $\mathcal{C}_0 \gets \{1, 2, \ldots, n\}$
\State \textbf{// Initialize with most representative sample}
\State $s_1 \gets \underset{i \in \mathcal{C}_0}{\arg\min} \left[ \frac{1}{n-1} \sum_{j \in \mathcal{C}_0 \setminus \{i\}} d(\mathbf{h}_i, \mathbf{h}_j) \right]$
\State $\mathcal{S}_1 \gets \{s_1\}$, $\mathcal{C}_1 \gets \mathcal{C}_0 \setminus \{s_1\}$
\For{$t = 2$ to $k$}
    \State \textbf{// Compute diversity scores}
    \For{$c \in \mathcal{C}_{t-1}$}
        \State $\mathcal{D}_{\text{ref}}(c) \gets \frac{1}{|\mathcal{S}_{t-1}|} \sum_{s \in \mathcal{S}_{t-1}} d(\mathbf{h}_c, \mathbf{h}_s)$
        \State $\mathcal{D}_{\text{query}}(c) \gets \frac{1}{|\mathcal{C}_{t-1}| - 1} \sum_{j \in \mathcal{C}_{t-1} \setminus \{c\}} d(\mathbf{h}_c, \mathbf{h}_j)$
    \EndFor
    \State \textbf{// Diversity-Representativeness score}
    \For{$c \in \mathcal{C}_{t-1}$}
        \State $r_1(c) \gets$ rank of $\mathcal{D}_{\text{ref}}(c)$ in descending order
        \State $r_2(c) \gets$ rank of $\mathcal{D}_{\text{query}}(c)$ in ascending order
        \State $\rho(c) \gets \frac{r_1(c) + r_2(c)}{2}$
    \EndFor
    \State \textbf{// Select optimal candidate}
    \State $s_t \gets \underset{c \in \mathcal{C}_{t-1}}{\arg\min} \, \rho(c)$
    \State $\mathcal{S}_t \gets \mathcal{S}_{t-1} \cup \{s_t\}$, $\mathcal{C}_t \gets \mathcal{C}_{t-1} \setminus \{s_t\}$
\EndFor
\State \Return $\mathcal{S}_k$
\end{algorithmic}
\end{algorithm}

\section{Reference Samples Selection}\label{sec:ref_sample}

In practice, users typically provide a set of representative reference image pairs that share similar visual characteristics with the images they intend to retouch. To simulate this behavior in our experiments and ensure that the selected references contain information relevant to the test images, we adopt a color–tone distribution–guided sampling strategy to partition each benchmark into reference and test splits.  

Specifically, we represent each image using histogram features extracted in the CIE–Lab color space. For each input image $x_i$, we compute a 1D histogram of the $L$ channel with 16 bins to capture the tonal distribution, and a 2D joint histogram of the \textit{ab} chromatic channels with $16 \times 16$ bins (256 bins in total) to model color variation. These are concatenated to form a unified color–tone feature vector $h_i \in \mathbb{R}^{272}$:  

\begin{equation}
h_i = [H_L(x_i); H_{ab}(x_i)].
\end{equation}

We then apply the sampling strategy described in~\Cref{alg:diverse_ref_sampling} to select references that are both representative and diverse. At each iteration, every candidate image is evaluated based on two complementary criteria: (i) its average distance to the already selected references, which encourages diversity, and (ii) its average distance to the remaining unselected candidates, which promotes representativeness. These two scores are combined using a simple rank-based fusion to obtain a single selection score for each image. The process is repeated in a greedy manner until the desired number of reference samples is obtained. The Chi-square distance is used to measure similarity between histogram features, as it effectively captures perceptual differences in color and tone distributions.  

We employ this reference sampling procedure across all three benchmarks. The same sampled references are consistently used for evaluating both baseline methods and our proposed model to ensure fairness. For each benchmark, we first sample the maximum number of references: 4, 6, and 100 pairs for VCIRB, PPR10K-groups, and the MIT-FiveK/PPR10K datasets, respectively and then derive smaller subsets by selecting the first $k$ references for experiments with fewer exemplars.

\section{Model Architecture Details} 

\paragraph{Overall Architecture.}
Our full pipeline consists of two main components that work together to achieve personalized image retouching: 
(1) an \emph{Asymmetric Auto-Encoder} that extracts a content-disentangled retouching style latent from paired images $(x, y)$, and 
(2) a \emph{retrieval-augmented retouching (RAR)} module that adaptively combines multiple reference retouching style latents to synthesize a query-specific retouching style. 
Given a user’s reference set $\mathcal{P}=\{(x_i, y_i)\}_{i=1}^K$, the asymmetric auto-encoder produces per-pair retouching style latents $\{z_i\}$, and RAR retrieves the most relevant references for a query image $x_{\mathrm{q}}$ based on content similarity. 
The aggregated retouching style latent $z_{\mathrm{q}}$ is finally decoded by a conditional MLP decoder to produce the retouched output $\hat{y}_{\mathrm{q}}$. 
The system is trained end-to-end for style encoding and decoding.

\paragraph{Asymmetric Style Auto-Encoder.}
Given a source image $x$ and its retouched counterpart $y$, the encoder $E_\theta(x, y)$ extracts a retouching style latent $z \in \mathbb{R}^{2048}$ describing the transformation applied in the pair.
We employ a Siamese SigLIPv2 backbone $\mathcal{B}$ (``google/siglip2-so400m-patch14-384''), operating at $384\times384$ resolution. 
Each attention block in $\mathcal{B}$ is augmented with LoRA adapters inserted into all linear layers within the self-attention module; only these LoRA weights are trainable, while base SigLIPv2 parameters remain frozen.

For each image, $\mathcal{B}$ produces a $1152$-dimensional pooled embedding. 
The embeddings from $x$ and $y$ are concatenated:
\begin{equation}
    h = [\,h_x ; h_y\,] \in \mathbb{R}^{1152 \times 2},
\end{equation}
and then projected through a linear layer to form the retouching style latent:
\begin{equation}
    z = W_{\text{proj}} h \in \mathbb{R}^{2048}.
\end{equation}

\paragraph{Conditional MLP Decoder.}
The decoder $D_\phi(x, z)$ reconstructs the retouched target using a pixel-wise conditional MLP.  
Given the original input image $x \in \mathbb{R}^{3\times H\times W}$ and its associated retouching style latent $z$, the pixels are reshaped to $(HW,3)$ and passed through an input linear layer $\mathbb{R}^3 \to \mathbb{R}^{128}$, followed by three hidden layers:
\begin{equation}
    128 \rightarrow 256 \rightarrow 512 \rightarrow 3.
\end{equation}
Each layer incorporates additive style conditioning:
\begin{equation}
    h_k = \mathrm{LayerNorm}(h_k) + W_k(z),
\end{equation}
with $W_k$ being a learned projection of $z$ for block $k$.  
ReLU activations follow intermediate layers, while a sigmoid is applied at the output.  
Thus, the decoder functions as a resolution-agnostic, style-conditioned pixel operator:
\begin{equation}
    \hat{y} = D_\phi(x, z).
\end{equation}

\paragraph{Retrieval-augmented retouching (RAR).}
To enable multi-reference personalized retouching, we introduce RAR, which aggregates style information from reference samples most relevant to the query image.  
Given a query image $x_{\mathrm{q}}$ and user references $(x_i, y_i)$ with retouching style latents $z_i$, we extract content embeddings using the SigLIPv2 encoder with LoRA disabled:
\begin{equation}
    c_{\mathrm{q}} = E_{\text{content}}(x_{\mathrm{q}}),
    \qquad
    c_i = E_{\text{content}}(x_i).
\end{equation}

The retouching style latents are obtained from the asymmetric encoder:
\begin{equation}
    z_i = E_\theta(x_i, y_i).
\end{equation}

RAR performs soft $k$-nearest-neighbor retrieval on the content embeddings.  
Let $s_i$ be the cosine similarity between $c_{\mathrm{q}}$ and $c_i$.  
We compute softmax reweighting with temperature $\tau=0.1$:
\begin{equation}
    \alpha_i = \frac{\exp\!\left(\frac{s_i}{\tau}\right)}
                    {\sum_{j \in \mathcal{N}_K}\exp\!\left(\frac{s_j}{\tau}\right)},
\end{equation}
and aggregate the corresponding retouching style latents:
\begin{equation}
    z_{\mathrm{q}} = \sum_{i \in \mathcal{N}_K} \alpha_i \, z_i.
\end{equation}

Finally, the personalized retouching output is generated by the conditional decoder:
\begin{equation}
    \hat{y}_{\mathrm{q}} = D_\phi(x_{\mathrm{q}}, z_{\mathrm{q}}).
\end{equation}

\paragraph{Final Formulation.}
Summarizing the entire inference pipeline:
\begin{equation}
    z_i = E_\theta(x_i, y_i), 
    c_i = E_{\text{content}}(x_i), 
    c_{\mathrm{q}} = E_{\text{content}}(x_{\mathrm{q}}),
\end{equation}
\begin{equation}
    z_{\mathrm{q}} = \mathrm{KNN}(c_{\mathrm{q}}, \{c_i\}, \{z_i\}),
\end{equation}
\begin{equation}
    \hat{y}_{\mathrm{q}} = D_\phi(x_{\mathrm{q}}, z_{\mathrm{q}}).
\end{equation}

\section{Inference Latency}
\label{sec:latency}

We evaluate the inference efficiency of \textbf{RefRetouch} under different input resolutions, decoder designs, and numbers of reference images. All measurements are conducted on a workstation equipped with an NVIDIA A100-80GB GPU with a 128-core Xeon CPU.

\paragraph{MLP decoder vs. 3D-LUT decoder.}
The default RefRetouch decoder is a conditional MLP that operates independently on each RGB pixel. For a fixed retouching style latent, this decoder defines a deterministic color mapping from the input RGB space to the output RGB space. Therefore, the pixel-wise MLP can be converted into a 3D-LUT by densely evaluating the decoder on a regular RGB grid and applying trilinear interpolation during inference. This replaces repeated per-pixel MLP evaluation with an efficient table lookup, which is particularly beneficial at higher resolutions.

To verify that this conversion does not degrade retouching quality, we compare the original MLP decoder with the 3D-LUT decoder in~\Cref{tab:refretouch_decoder_quality}. The results show that the 3D-LUT decoder achieves comparable PSNR to the MLP decoder across both PPR10K-Groups and MIT-FiveK, indicating that the LUT preserves the learned retouching transformation.

We further compare the decoding efficiency of the two decoder designs in~\Cref{tab:decoder_varying_resolution}. Since the MLP decoder is evaluated for every pixel, its latency and memory usage increase substantially with resolution. In contrast, the 3D-LUT decoder performs retouching through table lookup and interpolation, resulting in nearly constant latency across resolutions and significantly lower memory consumption.

\begin{table*}[t]
\centering
\caption{Decoder-only efficiency comparison under different output resolutions. ``Lat.'' denotes inference latency measured in milliseconds at batch size 1, ``Thr.'' denotes throughput measured in images per second at batch size 4, and ``VRAM'' denotes peak GPU memory usage measured in MB.}
\label{tab:decoder_varying_resolution}
\small
\setlength{\tabcolsep}{4.0pt}
\renewcommand{\arraystretch}{1.10}
\begin{tabular}{@{}lccc ccc ccc@{}}
\toprule
\multirow{2}{*}{\textbf{Method}}
& \multicolumn{3}{c}{$256 \times 256$}
& \multicolumn{3}{c}{$512 \times 512$}
& \multicolumn{3}{c}{$1024 \times 1024$} \\
\cmidrule(lr){2-4} \cmidrule(lr){5-7} \cmidrule(lr){8-10}
& \textbf{Lat.} & \textbf{Thr.} & \textbf{VRAM}
& \textbf{Lat.} & \textbf{Thr.} & \textbf{VRAM}
& \textbf{Lat.} & \textbf{Thr.} & \textbf{VRAM} \\
\midrule
3D-LUT & 1.66 & 653.88 & 1944.8 & 1.67  & 648.26 & 1949.3 & 1.72  & 622.67 & 1983.3 \\
MLP    & 2.84 & 366.95 & 2088.0 & 10.85 & 92.61  & 3052.5 & 43.21 & 23.01  & 6926.5 \\
\bottomrule
\end{tabular}
\end{table*}

\begin{table}[t]
\centering
\caption{Quality comparison between the MLP decoder and the 3D-LUT decoder in RefRetouch. Results are reported in PSNR (dB), where higher is better.}
\label{tab:refretouch_decoder_quality}
\small
\setlength{\tabcolsep}{4pt}
\renewcommand{\arraystretch}{1.1}
\resizebox{\linewidth}{!}{%
\begin{tabular}{lcccccc}
\toprule
\multirow{2}{*}{\textbf{Method}}
& \multicolumn{3}{c}{\textbf{PPR10K-Groups}}
& \multicolumn{3}{c}{\textbf{MIT-FiveK}} \\
\cmidrule(lr){2-4} \cmidrule(lr){5-7}
& \textbf{1} & \textbf{3} & \textbf{6}
& \textbf{20} & \textbf{50} & \textbf{100} \\
\midrule
RefRetouch + MLP    & 22.51 & 23.92 & 25.27 & 22.95 & 23.19 & 23.33 \\
RefRetouch + 3D-LUT & 22.51 & 23.92 & 25.28 & 22.97 & 23.21 & 23.34 \\
\bottomrule
\end{tabular}
}
\end{table}

\paragraph{Scalability with the number of references.}
We also evaluate the inference cost under varying numbers of reference images, denoted as $N_{\mathrm{ref}}$. As shown in~\Cref{tab:varying_num_references}, the latency of RefRetouch increases with $N_{\mathrm{ref}}$ because the reference inputs are encoded for retrieval. However, for a fixed user reference set, the reference content embeddings can be precomputed and cached. Thus, repeated inference for new query images only requires query encoding, retrieval over cached features, style aggregation, and decoding.

\begin{table*}[t]
\centering
\caption{Inference latency and peak memory usage under different numbers of reference images. Latency is reported in milliseconds and memory is reported in MB. }
\label{tab:varying_num_references}
\scriptsize
\resizebox{\textwidth}{!}{%
\begin{tabular}{lcc cc cc cc cc}
\toprule
\multirow{2}{*}{\textbf{Method}}
& \multicolumn{2}{c}{$N_{\mathrm{ref}}=1$}
& \multicolumn{2}{c}{$N_{\mathrm{ref}}=3$}
& \multicolumn{2}{c}{$N_{\mathrm{ref}}=5$}
& \multicolumn{2}{c}{$N_{\mathrm{ref}}=10$}
& \multicolumn{2}{c}{$N_{\mathrm{ref}}=50$} \\
\cmidrule(lr){2-3} \cmidrule(lr){4-5} \cmidrule(lr){6-7} \cmidrule(lr){8-9} \cmidrule(lr){10-11}
& \textbf{Lat.} & \textbf{VRAM}
& \textbf{Lat.} & \textbf{VRAM}
& \textbf{Lat.} & \textbf{VRAM}
& \textbf{Lat.} & \textbf{VRAM}
& \textbf{Lat.} & \textbf{VRAM} \\
\midrule
SigLIPv2     & 166.6 & 3028.3 & 712.3 & 3040.4 & 1091.9 & 3102.4 & 1899.9 & 3178.5 & 9290.6 & 4421.5 \\
DINOv2        & 141.4 & 1996.4 & 572.5 & 2044.4 & 676.0  & 2106.5 & 806.8  & 2182.5 & 2791.8 & 2783.1 \\
StarEnhancer  & 7.8   & 214.5  & 8.7   & 322.0  & 12.7   & 432.0  & 21.7   & 700.0  & 97.4   & 1996.0 \\
MSM           & 9.1   & 833.1  & 9.0   & 837.6  & 9.3    & 840.0  & 10.8   & 846.6  & 26.3   & 998.5 \\
\bottomrule
\end{tabular}
}
\end{table*}

The results indicate that the 3D-LUT decoder provides substantial efficiency gains over the pixel-wise MLP decoder while preserving reconstruction quality. The remaining reference-dependent cost comes from retrieval encoding, which can be redcued through caching for a fixed reference set.

\section{User Study Details}

\begin{figure*}[t]
    \centering
    \begin{subfigure}{0.48\linewidth}
        \centering
        \includegraphics[width=\linewidth]{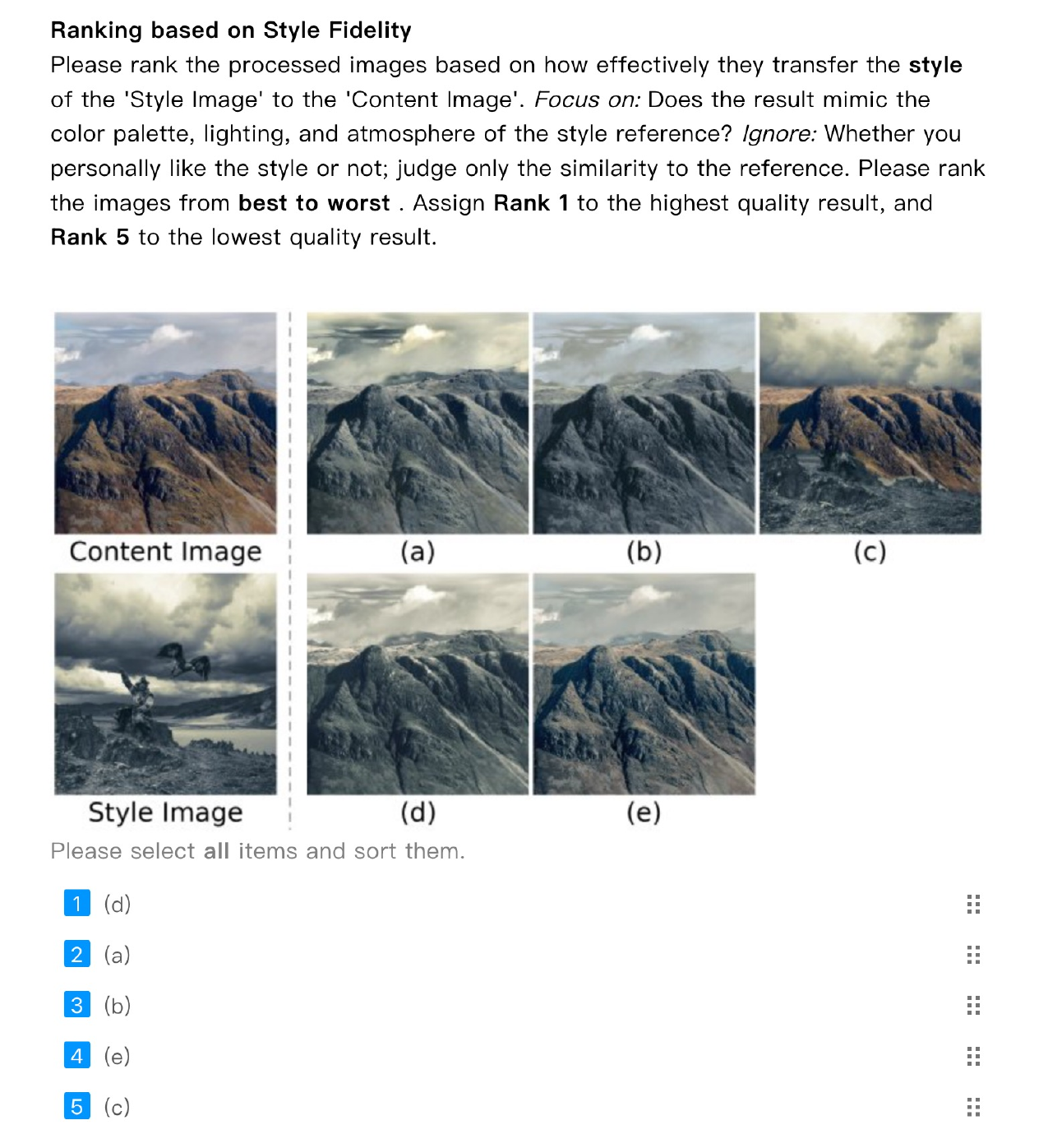}
        \caption{Style fidelity}
        \label{fig:style_transfer_study}
    \end{subfigure}
    \hfill
    \begin{subfigure}{0.48\linewidth}
        \centering 
        \includegraphics[width=\linewidth]{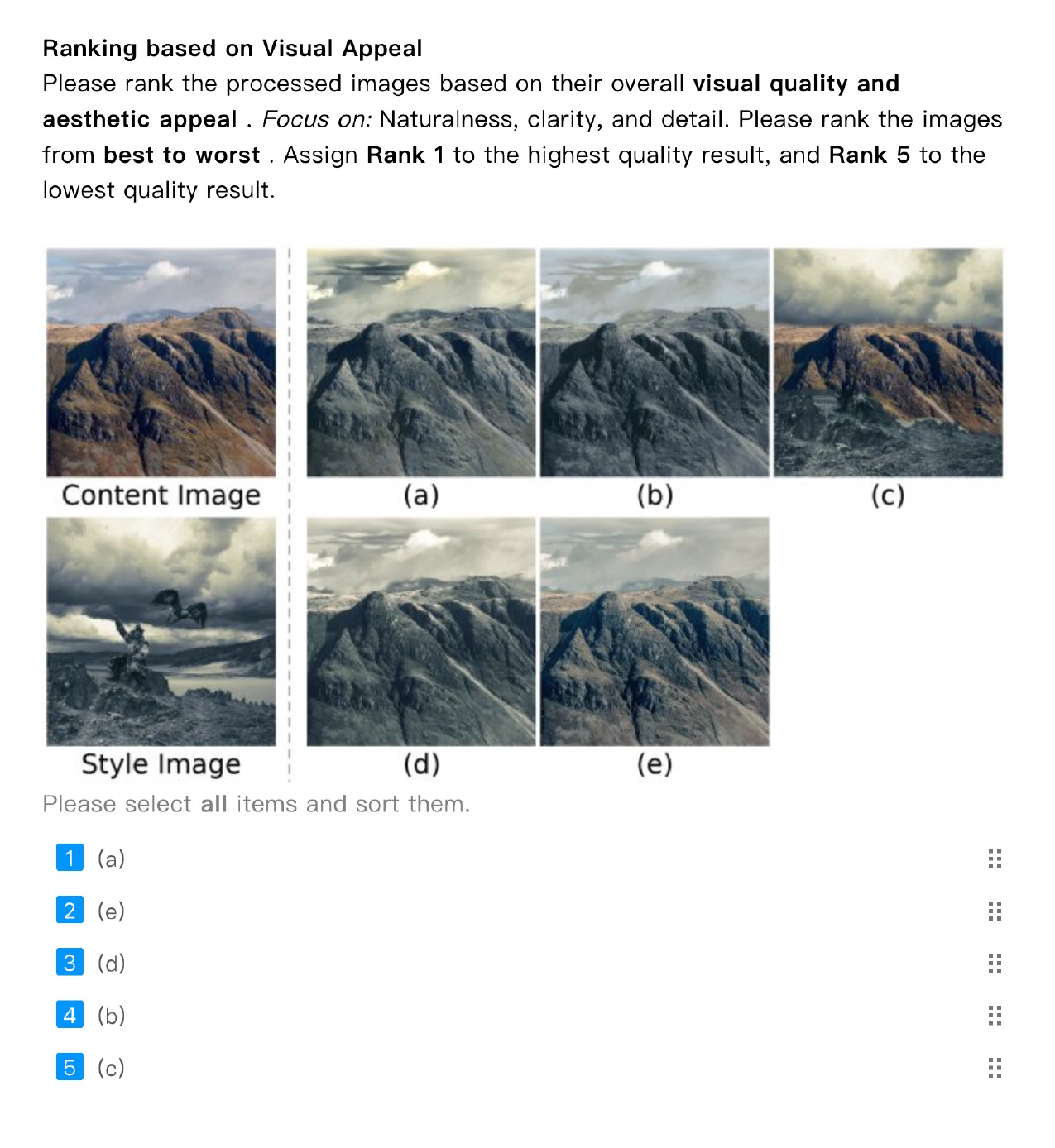}
        \caption{Visual appeal}
        \label{fig:visual_appeal_study}
    \end{subfigure}
    \caption{An example interface of our user study. Each set of results is assessed under two criteria, \emph{style transfer fidelity} and \emph{visual appeal}, each accompanied by clear task instructions. Participants rank the results by dragging the choice labels according to the performance of each method under the corresponding criterion. The labels assigned to the compared approaches are fully randomized to mitigate bias.}
    \label{fig:user_study_interface}
\end{figure*}

Evaluating photorealistic style transfer is inherently challenging because, unlike other personalized image retouching tasks, it lacks a ground-truth target against which model outputs can be directly compared. Although prior work has proposed quantitative metrics for this task~\cite{ke2023neural}, which we also reported in Table~2 of the main paper, they do not fully capture the model's ability to perform faithful visual transfer. We therefore conduct a user study that evaluates the results from the user’s perspective along two dimensions: \emph{style transfer fidelity} and \emph{visual appeal}, as shown in~\Cref{fig:user_study_interface}. We compare our method with the four approaches that obtain the highest style scores in Table~2, namely D-LUT, DeepPreset, PhotoWCT2, and NanoBanana. In total, we select 40 photorealistic style transfer samples from the results on VCIRB, and each participant is assigned 10 randomly chosen evaluation questions to ensure broad coverage of the sample set. 

\section{Retrieval Implementation Details}
\label{sec:retrieval_implementation}

We provide the implementation details for the retrieval analysis-based feature analysis in the main paper. For each evaluated encoder, we first extract embeddings for all images in the selected VCIRB split and use them to construct the retrieval candidate pool. For our adapted representation, we use the LoRA-adapted SigLIPv2 backbone from the Siamese style encoder and take its pooled feature representation before the final projection layer. For comparison, we also extract embeddings from the frozen SigLIPv2 encoder. All embeddings are normalized before retrieval.

During retrieval analysis, all images in the selected VCIRB split are used as candidate references, while candidates from the same content group as the query are excluded. This prevents near-duplicate or highly similar semantic content from dominating the retrieved results and makes the comparison focus more clearly on cross-content similarity. Therefore, the retrieved neighbors reflect what each embedding space considers similar across different content groups.

\begin{figure*}[!t]
    \centering
    \includegraphics[width=0.98\textwidth,height=0.68\textheight,keepaspectratio]{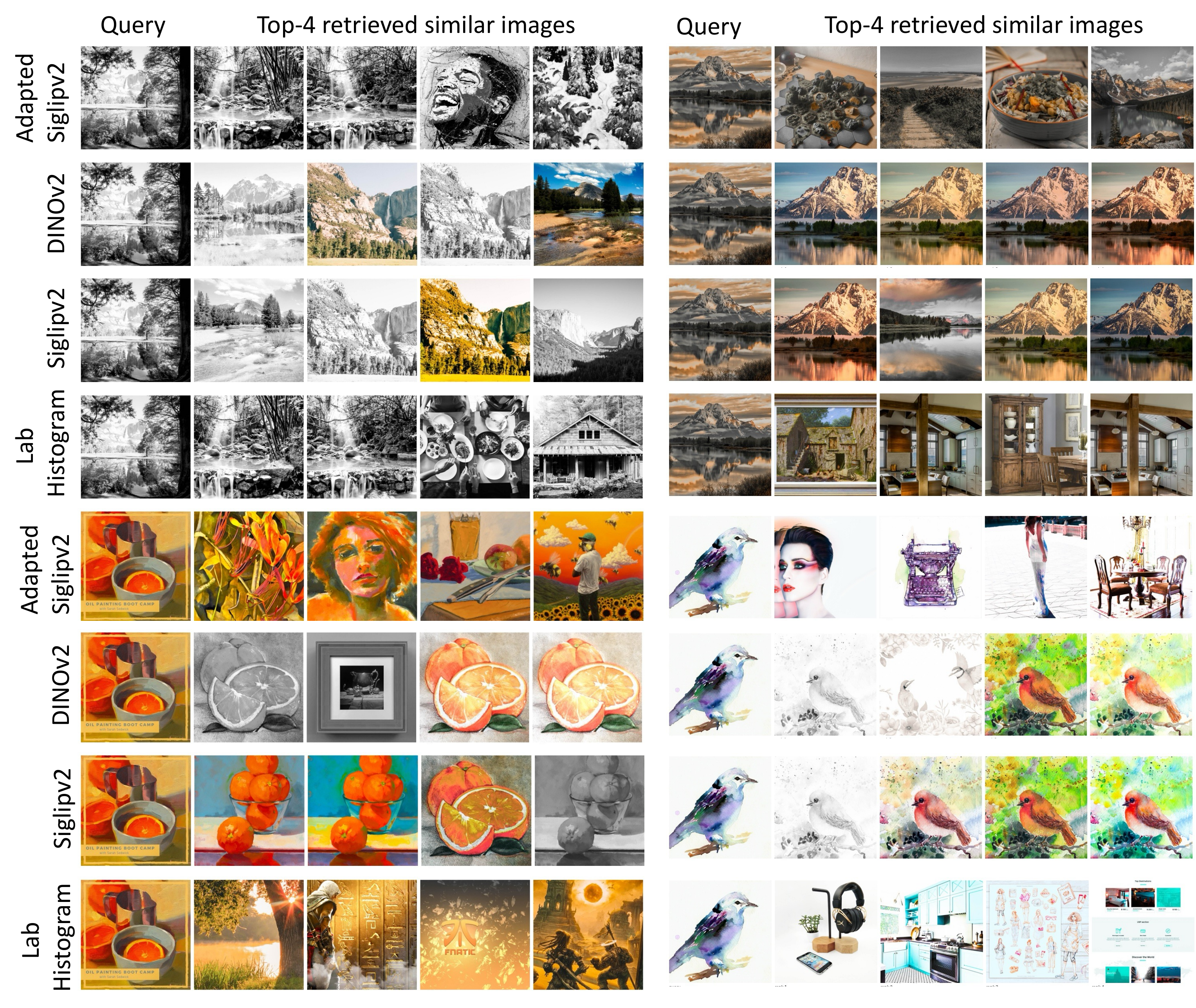}
    \caption{Additional retrieval comparison across different feature embeddings.}
    \label{fig:retrieval2}
\end{figure*}

\Cref{fig:retrieval2} provides additional retrieval results with two additional encoders: DINOv2 and Histogram-Lab. DINOv2 is used as an additional frozen semantic representation baseline, allowing us to compare the adapted SigLIPv2 features against a strong generic visual representation. Histogram-Lab is used as a low-level photometric baseline. Specifically, each image is represented by a CIELAB histogram descriptor consisting of a 16-bin histogram over the luminance channel and a joint $16 \times 16$ histogram over the two chromatic channels. The two histograms are independently normalized and concatenated into a 272-dimensional feature vector. This baseline captures global color and tone statistics without relying on learned semantic or retouching-specific representations.

This extended comparison allows us to examine whether the LoRA-adapted SigLIPv2 encoder captures retouching-relevant similarity after asymmetric auto-encoding training. The adapted features retrieve images with more consistent color and tone characteristics across different content groups, while frozen semantic encoders are more strongly influenced by object- or scene-level similarity. Compared with the Histogram-Lab baseline, the adapted features also provide a learned representation of photometric similarity rather than relying solely on global color statistics, supporting their use as the content encoder in retrieval-augmented retouching.

\makeatletter
\setlength{\@dblfptop}{0pt}
\makeatother

\section{Additional Qualitative Results}

\begin{figure*}
    \centering
    \includegraphics[width=0.96\textwidth]{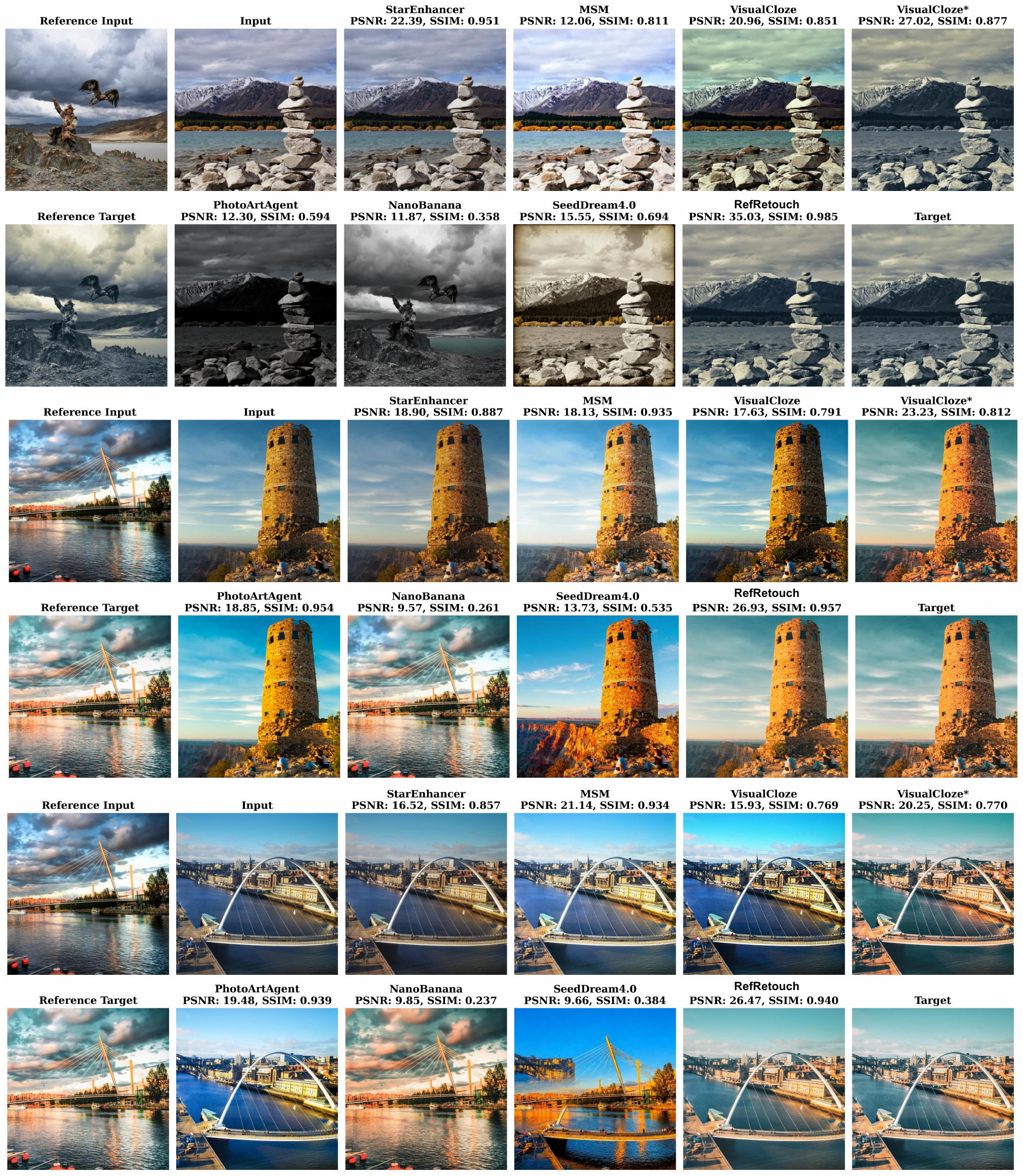}
    \caption{VCIRB single reference personalized image retouching qualitative result comparisons.}
    \label{fig:single_ref}
\end{figure*}

\begin{figure*}
    \centering
    \includegraphics[width=0.78\textwidth]{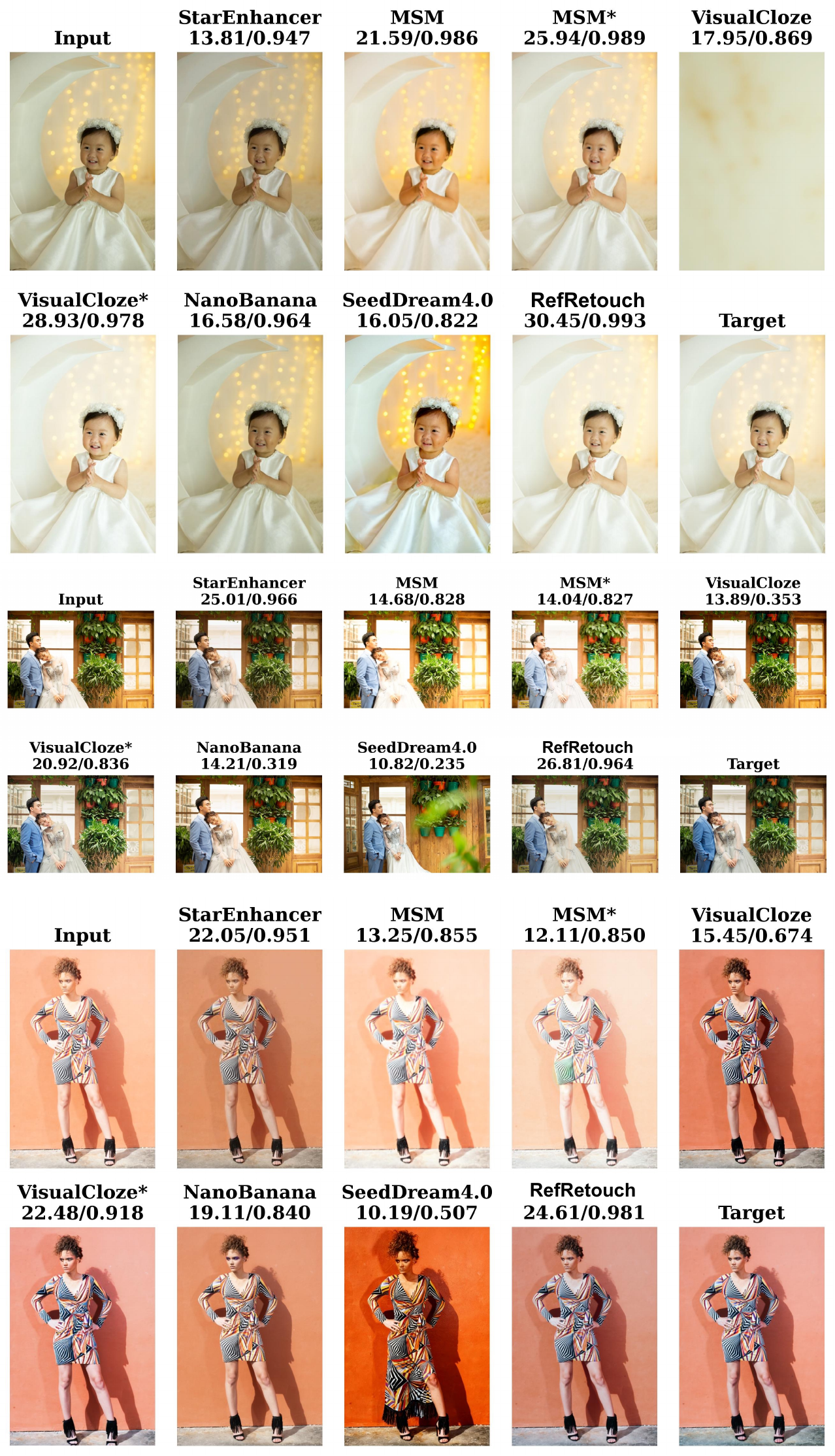}
    \caption{PPR10k-groups single reference personalized image retouching qualitative result comparisons.}
    \label{fig:ppr10k_grps_single_ref}
\end{figure*}

\begin{figure*}
    \centering
    \includegraphics[width=0.78\textwidth]{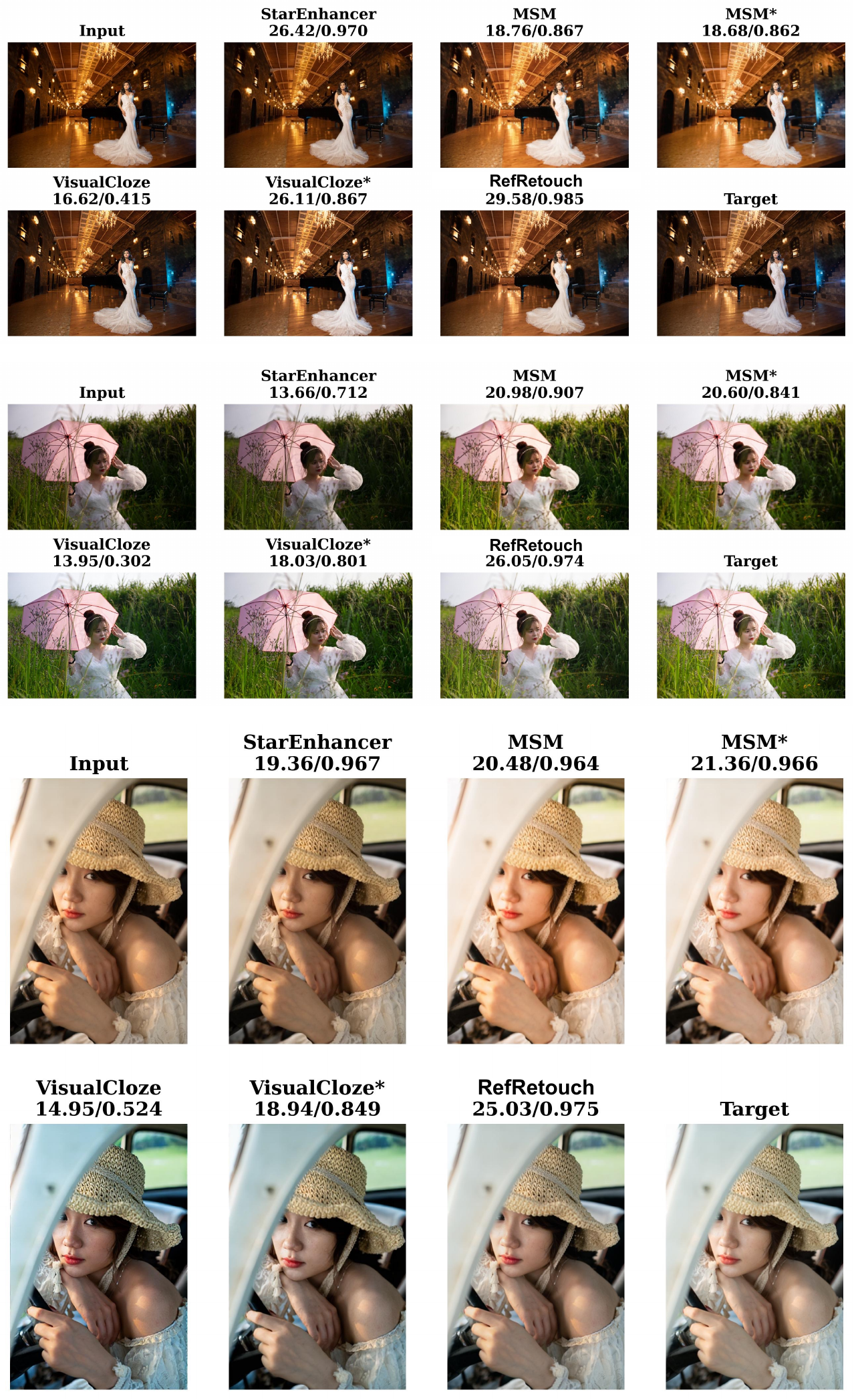}
    \caption{PPR10k-groups three references personalized image retouching qualitative result comparisons.}
    \label{fig:ppr10k_grps_three_ref}
\end{figure*}

\begin{figure*}[t]
    \centering
    \includegraphics[width=0.78\textwidth]{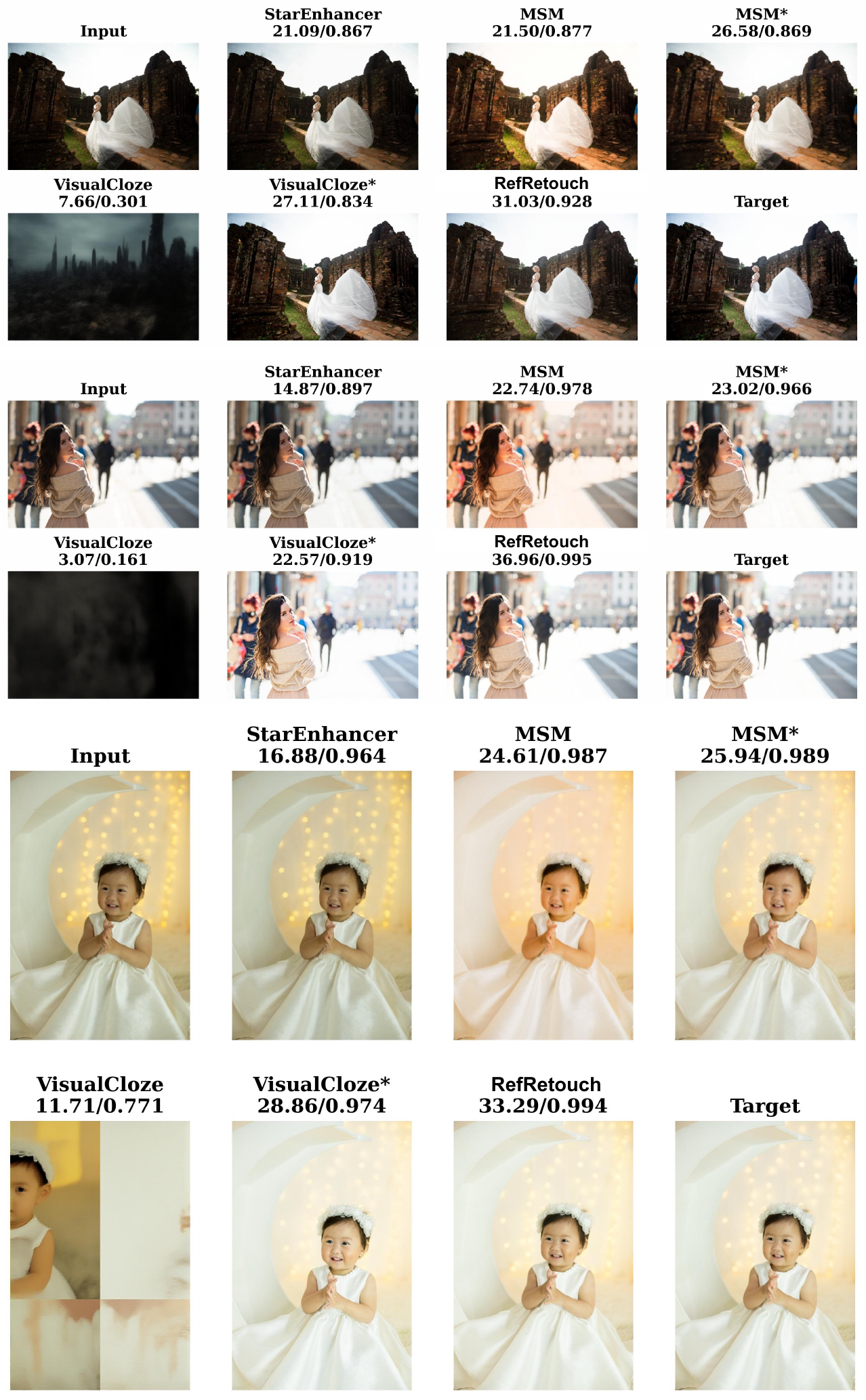}
    \caption{PPR10k-groups six references personalized image retouching qualitative result comparisons.}
    \label{fig:ppr10k_grps_six_ref}
\end{figure*}

\begin{figure*}[t]
    \centering
    \includegraphics[width=\textwidth,height=0.98\textheight,keepaspectratio]{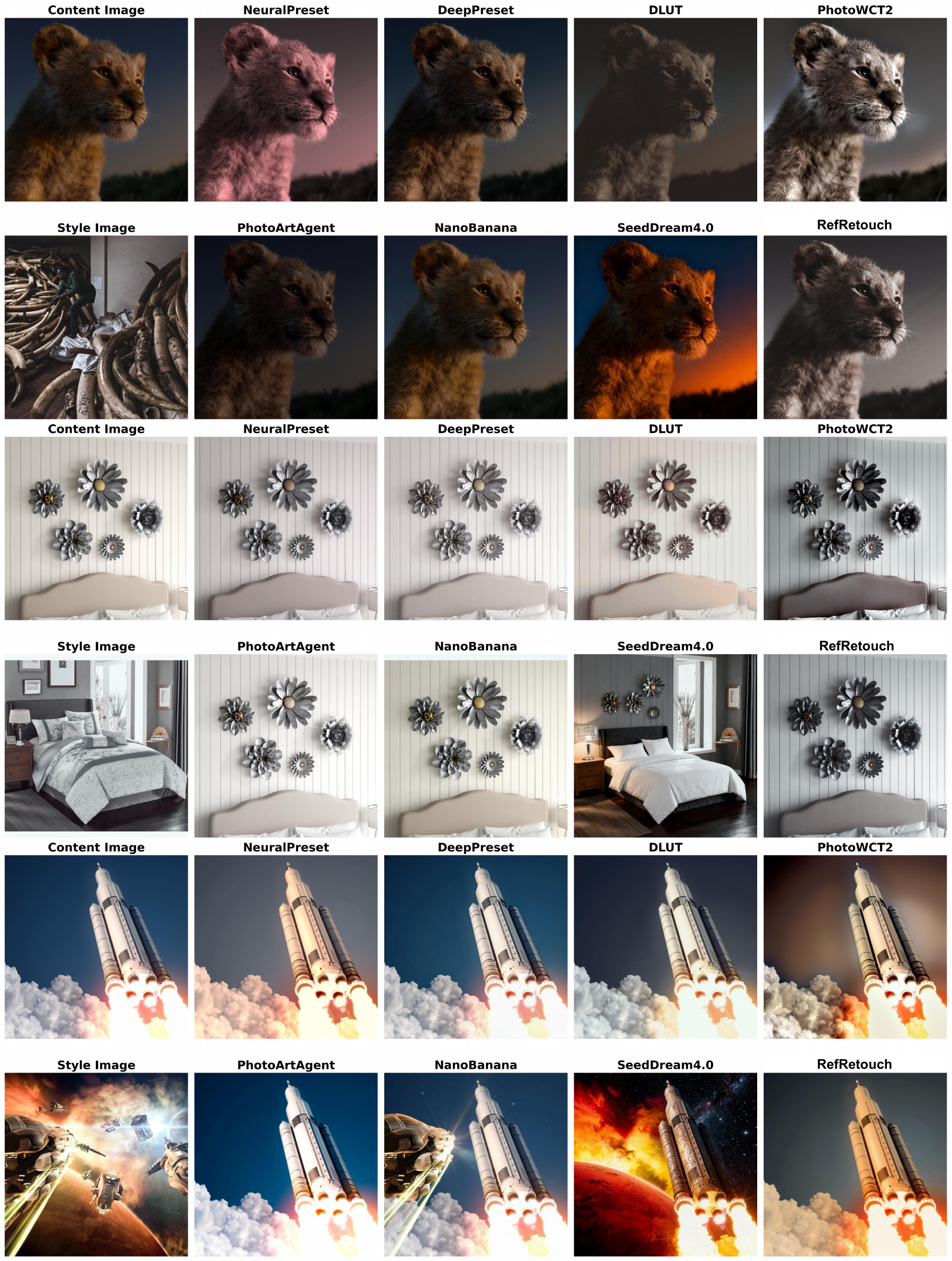}
    \caption{Photorealistic style transfer qualitative result comparisons.}
    \label{fig:target_only}
\end{figure*}

\begin{figure*}[t]
    \centering
    \includegraphics[width=0.94\textwidth]{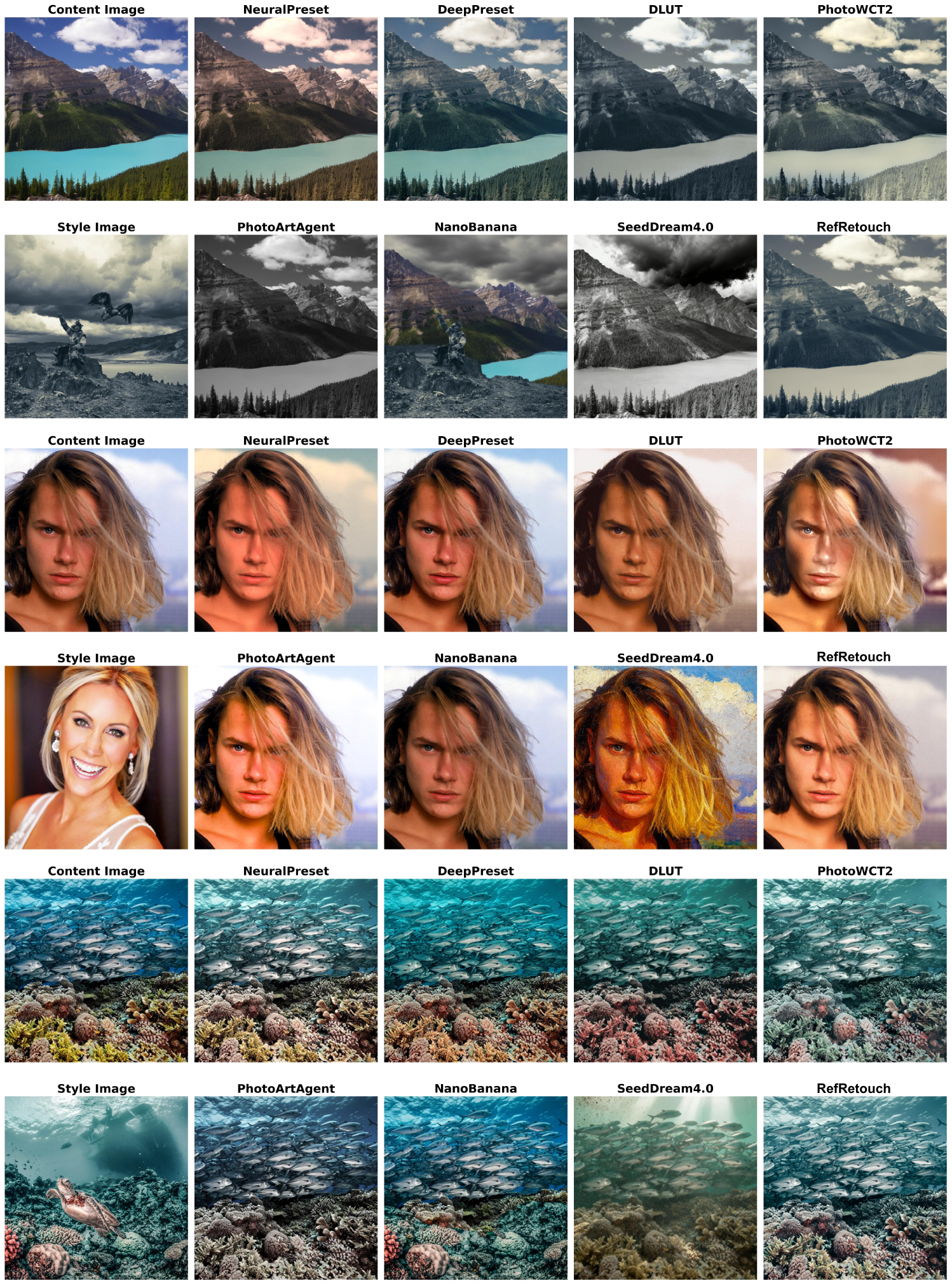}
    \caption{Photorealistic style transfer qualitative result comparisons.}
    \label{fig:target_only2}
\end{figure*}

We provide additional qualitative results demonstrating the visual effectiveness and generalization capability of our method across different personalization scenarios. \Cref{fig:single_ref} shows personalization results from a single reference pair, while~\Cref{fig:ppr10k_grps_single_ref}, \Cref{fig:ppr10k_grps_three_ref}, and~\Cref{fig:ppr10k_grps_six_ref} illustrate our model’s performance on the PPR10K-groups benchmark with one, three, and six reference exemplars, respectively. Furthermore, \Cref{fig:target_only} and \Cref{fig:target_only2} present additional photorealistic style transfer comparisons across diverse content and style references. 

\end{document}